# Exceptional deficiency of non-Hermitian systems

Zhen Li[1], Rundong Cai[1], Xulong Wang[1], Kenji Shimomura[2], Congwei Lu[1], Zhesen Yang[3*], Masatoshi Sato[2*], Guancong Ma[1,4*]

[1]*Department of Physics, Hong Kong Baptist University, Kowloon Tong, Hong Kong, China.*

[2]*Center for Gravitational Physics and Quantum Information, Yukawa Institute for Theoretical Physics, Kyoto University, Kyoto 606-8502, Japan.*

[3]*Department of Physics, Xiamen University, Xiamen 361005, Fujian, China.*

[4]*Shenzhen Institute for Research and Continuing Education, Hong Kong Baptist University, Shenzhen 518000, China.*

**Exceptional points are non-Hermitian singularities associated with the coalescence of individual eigenvectors accompanied by the degeneracy of their complex energies. Since their discovery, exceptional points have attracted much interest and enabled numerous advanced applications including sensing. However, accessing exceptional points generally requires delicate parameter tuning, and any related phenomena is intrinsically restricted to a very narrow bandwidth. Here, we report a generalization to the concept of exceptional points called exceptional deficiency that features the complete coalescence of entire eigenspaces with identical but arbitrarily large dimensions and the coincidence of entire spectral continua. We find that exceptional deficiency can induce anomalous absence and presence of non-Hermitian skin effect that transcends the established topological bulk-edge correspondence, resulting in unexpected synergistic skin-propagative dynamics. These phenomena are experimentally observed using active mechanical lattices. We further explore how exceptional deficiency offers a route for reliable and flexible control of localization and propagation, and enables a high-sensitivity broadband sensor. The experimental demonstration of exceptional deficiency provides a new perspective on non-Hermitian physics and may impact related applications such as sensing, modal control and lasing.**

*emails: yangzs@xmu.edu.cn (Z.Y.), msato@yukawa.kyoto-u.ac.jp (M.S.), phgcma@hkbu.edu.hk (G.M.)

Exceptional points (EPs) are unique singularities in the complex spectrum of non-Hermitian systems[1,2], which are open systems governed by both internal degrees of freedom and energy (or particle) exchange with the external. At an EP, the eigenvectors of two states (or more, in cases of higher-order EPs) coalesce, accompanied by the degeneracy of the corresponding eigenvalues. The Hilbert space resultantly loses a dimension and becomes defective[3,4], leading to a Jordan canonical form of the Hamiltonian. The investigations of EPs have led to a multitude of revolutionary physics[5–8] and diverse phenomena with rich application potentials[9–12]. Yet, because EP formations only involve an $\mathcal{O}(1)$ number of states isolated in the spectrum, related phenomena and functionalities are intrinsically limited to a very narrow bandwidth and require precise control to access.

In this work, we generalize the concept of EP by considering large non-Hermitian Hamiltonians with a block-triangular form, which was previously studied for hierarchical generation of higher-order EPs[13–15]. We discover that two high-dimensional eigenspaces, rather than individual eigenvectors, can align completely. (In this work, high-dimensional refers to the dimension of the eigenspace instead of spatial dimensions.) We denote this as exceptional deficiency (ED). At ED, the spectra of two coalescing eigenspaces coincide, and the Hilbert space is $\mathcal{O}(N)$ defective, with defective states emerging over a spectral continuum instead of at isolated spectral points. The properties of the ED enable non-Hermitian dynamics in systems consisting of two large eigenspaces of equal dimensions, one is Hermitian, the other is non-Hermitian. The latter is realized as a lattice under non-Hermitian skin effect (NHSE), a localization mechanism that turns states in continuum bands into skin modes clinging to an open boundary[16–19]. As a crucial consequence of the ED, the celebrated topological bulk-edge correspondence of NHSE[20] is broken, leading to unprecedented non-Hermitian dynamic effects characterized by the synergy of NHSE and propagation. The condition of ED also unveils a universal and convenient route for the reliable and flexible control of localization and propagation in non-Hermitian systems, and enables broadband, high-sensitivity detectors with a large dynamic range. Our work opens multiple horizons for non-Hermitian physics and applications across many realms.

**Generalization of EP**

We begin with a simple two-level Hamiltonian, $\mathbf{H} = \begin{pmatrix} h_A & \kappa \\ 0 & h_B \end{pmatrix}$, which is non-Hermitian when $\kappa \neq 0$. Figure 1a,b plots the eigenvalues and eigenvectors as functions of $\Delta h = |h_A - h_B|$. When $\Delta h$ reduces to zero, the eigenvalues become identical, and the eigenvectors are increasingly skewed and eventually aligned. At this point, $\mathbf{H}$ is a Jordan block, and an EP is reached[3,21].

Consider a generalization that replaces all entries in $\mathbf{H}$ with $N \times N$ square matrices

$$\mathbf{H} = \begin{pmatrix} \mathbf{h}_A & \boldsymbol{\kappa} \\ \mathbf{0} & \mathbf{h}_B \end{pmatrix}. \tag{1}$$

In the simple case where $\boldsymbol{\kappa} = \mathbf{0}$ and $\mathbf{h}_{A,B}$ are Hermitian, $\mathbf{H} = \mathbf{h}_A \oplus \mathbf{h}_B$ is spanned by the eigenvectors of $\mathbf{h}_A$ and $\mathbf{h}_B$, which form a complete set of orthonormal bases. However, when $\boldsymbol{\kappa} \neq \mathbf{0}$ and $\mathbf{h}_A = \mathbf{h}_B$, $\mathcal{E}(\mathbf{h}_B)$ coalesces with $\mathcal{E}(\mathbf{h}_A)$, where $\mathcal{E}(\cdot)$ denotes eigenspace: the Hilbert space of $\mathbf{H}$ is $\mathcal{O}(N)$ defective and half of the span of it is missing, as graphically depicted in Fig. 1c. We denote this situation as the ED. (Here, $\mathcal{E}(\mathbf{h}_{A,B})$ is formed by the eigenvectors of $\mathbf{H}$ with eigenvalues in the spectra $\eta(\mathbf{h}_{A,B})$ of $\mathbf{h}_{A,B}$. It does not refer to the original Hilbert space of $\mathbf{h}_{A,B}$.)

The ED hinges on the block-triangular form of $\mathbf{H}$ (Supplementary Section 1), which has an invariant subspace[22]. It can be proved that the necessary condition of the ED is $\eta(\mathbf{h}_A) = \eta(\mathbf{h}_B)$ (Methods). In other words, rather surprisingly, ED can appear even when $\mathbf{h}_A$ and $\mathbf{h}_B$ are different matrices. An intriguing example is shown in Fig. 1d, where $\mathbf{h}_A$ is Hermitian and $\mathbf{h}_B$ is non-Hermitian. The skewed $\mathcal{E}(\mathbf{h}_B)$ coalesces with the orthogonal $\mathcal{E}(\mathbf{h}_A)$. Remarkably, the highly defective Hilbert space at the ED remains an orthogonal linear space (Fig. 1d), which is in stark contrast with conventional EPs. On the contrary, if $\mathbf{h}_A$ and $\mathbf{h}_B$ are exchanged, the Hilbert space is skewed at the ED (Fig. 1e). The skewness is not controlled by $\boldsymbol{\kappa}$; it is a characteristic of the non-Hermitian block. These interesting characteristics are highly consequential in the following systems.

**ED and the anomalous absence and presence of NHSE**

Now we investigate the exotic physical phenomena induced by the ED. Two examples, which are one-dimensional double-chain lattices, are shown in Fig. 2a. In system-I, two Su-Schrieffer-Heeger chains, one Hermitian (denoted chain-A) and the other non-Hermitian with asymmetric intra-cell hopping (denoted chain-B), are one-way coupled by hopping $\kappa_1$ that hops only upwards (from chain-B to A). In system-II, everything is the same except that the one-way hopping $\kappa_2$ hops downwards (from chain-A to B). The Bloch Hamiltonians of the two systems are both block-triangular matrices

$$\boldsymbol{\mathcal{H}}_{\mathrm{I}}(k) = \begin{pmatrix} \boldsymbol{\mathcal{H}}_A(k) & \kappa_1 \mathbf{I}_2 \\ \mathbf{0} & \boldsymbol{\mathcal{H}}_B(k) \end{pmatrix}, \boldsymbol{\mathcal{H}}_{\mathrm{II}}(k) = \begin{pmatrix} \boldsymbol{\mathcal{H}}_A(k) & \mathbf{0} \\ \kappa_2 \mathbf{I}_2 & \boldsymbol{\mathcal{H}}_B(k) \end{pmatrix}. \tag{2}$$

Here, $\boldsymbol{\mathcal{H}}_A(k) = d_{Ax}\boldsymbol{\sigma}_x + d_{Ay}\boldsymbol{\sigma}_y$, with $d_{Ax} = v_1 + w\cos k$, $d_{Ay} = w\sin k$, where $\boldsymbol{\sigma}_{x,y}$ are the Pauli matrices, $\mathbf{I}_2$ is a $2 \times 2$ identity matrix; and $\boldsymbol{\mathcal{H}}_B(k) = d_{Bx}\boldsymbol{\sigma}_x + d_{By}\boldsymbol{\sigma}_y$, with $d_{Bx} = v_2 +$ $w\cos k$, $d_{By} = w\sin k + i\delta$. These two seemingly unassuming lattices have some rather surprising characteristics. First, since the characteristic equations of $\boldsymbol{\mathcal{H}}_{\mathrm{I,II}}$ can reduce to $\boldsymbol{\mathcal{H}}_A$ and $\boldsymbol{\mathcal{H}}_B$, the systems' spectra under a periodic boundary condition (PBC) are $\eta(\boldsymbol{\mathcal{H}}_{\mathrm{I,II}}) = \eta(\boldsymbol{\mathcal{H}}_A) \cup \eta(\boldsymbol{\mathcal{H}}_B)$, which are independent of the unidirectional inter-chain hopping $\kappa_{1,2}$. Therefore, the two systems have identical PBC spectra, which is the union of two distinct sub-spectra: a pair of loop spectra with identical nontrivial winding and a pair of real-valued spectra (Fig. 2b). And because $\eta(\boldsymbol{\mathcal{H}}_A) \neq \eta(\boldsymbol{\mathcal{H}}_B)$, ED does not appear in the PBC systems.

Let us now examine the open boundary condition (OBC) Hamiltonians, denoted $\boldsymbol{\hbar}_{\mathrm{I}}$ and $\boldsymbol{\hbar}_{\mathrm{II}}$, which also have $\eta(\boldsymbol{\hbar}_{\mathrm{I,II}}) = \eta(\boldsymbol{\hbar}_A) \cup \eta(\boldsymbol{\hbar}_B)$. Here the OBC spectra of both chains, $\eta(\boldsymbol{\hbar}_A)$ and $\eta(\boldsymbol{\hbar}_B)$, are real. When $v_1 = v_e = -\sqrt{(v_2-\delta)(v_2+\delta)}$, $\eta(\boldsymbol{\hbar}_A) = \eta(\boldsymbol{\hbar}_B)$ (Fig. 2d) and OBC Hamiltonians reach ED (Supplementary Section 2). In system-I, $\mathcal{E}(\boldsymbol{\hbar}_B)$ becomes defective. Examining the OBC eigenstates, we see that no state is affected by NHSE – all states are fully extended (Fig. 2e). For the same reason, $\mathcal{E}(\boldsymbol{\hbar}_A)$ is defective in system-II, and the OBC eigenstates thereof are entirely skin modes (Fig. 2f). These situations clearly deviate from the prevailing theories about the correspondence between NHSE and nontrivial PBC spectral topology[18,19,23] since system-I and II have the same PBC and OBC spectra but totally different localization properties. We remark that ED can appear in other, diverse double-chain systems (Supplementary Sections 3 and 4).

To gain more insights, we analyze the systems using the non-Bloch theory[24,19]. Note that the characteristic polynomials $f_{\mathrm{I,II}}(\beta, E) = \det[\boldsymbol{\mathcal{H}}_{\mathrm{I,II}}(\beta) - E\mathbf{I}_4] = 0$ have no dependence on $\kappa_{1,2}$, and is exactly factorizable as $f_{\mathrm{I,II}}(\beta, E) = f_A(\beta, E) f_B(\beta, E)$, where $\beta \coloneqq e^{i(k+ik')}$. Inserting the identical OBC spectra of the two systems, i.e., $E \in \eta(\boldsymbol{\hbar}_{\mathrm{I(II)}})$, into the characteristic polynomials, we can obtain the solution that consists of two parts: one is a unit circle $\beta_A = e^{ik}$ representing Bloch waves associated with the Hermitian chain-A, the other is the non-Bloch waves given by $\beta_B = \sqrt{(v_2-\delta)/(v_2+\delta)}\, e^{ik}$, which only relates to the non-Hermitian chain-B (Fig. 2c). From the OBC eigenstates shown in Fig. 2e (f), it is clear that system-I (II) is dominated by sub-generalized Brillouin zone $\beta_A$ ($\beta_B$), which corresponds to Bloch (non-Bloch) waves. This anomalous effect is rooted in the unique property of the ED, at which an entire eigenspace spanned by non-Bloch (Bloch) waves becomes defective.

We use the cosine similarity to quantify the behavior of the coalescing eigenspaces

$$\mathcal{C}\big(\mathcal{E}(\boldsymbol{\hbar}_A), \mathcal{E}(\boldsymbol{\hbar}_B)\big) = \frac{\langle \mathcal{E}(\boldsymbol{\hbar}_A), \mathcal{E}(\boldsymbol{\hbar}_B)\rangle_F}{\|\mathcal{E}(\boldsymbol{\hbar}_A)\|_F \, \|\mathcal{E}(\boldsymbol{\hbar}_B)\|_F}, \tag{3}$$

where $\langle\cdot,\cdot\rangle_F$ is the Frobenius inner product and $\|\cdot\|_F$ is the Frobenius norm (Methods). The cosine similarity is an extension of the vector inner product – it essentially projects a multi-dimensional linear space $\mathcal{E}(\boldsymbol{\hbar}_A)$ onto another $\mathcal{E}(\boldsymbol{\hbar}_B)$ to evaluate their similarity. The results are plotted in Fig. 3a (b) for system-I (system-II) as functions of $v_1$ and $\kappa_1$ ($\kappa_2$). It is seen that when $v_1 = v_e$ , ED is reached with an infinitesimally small $\kappa_{1(2)}$ and the two eigenspaces perfectly align, resulting in $\mathcal{C} = 1$. Even when the condition of ED is not exactly reached, i.e., $v_1$ slightly deviates from $v_e$ , $\mathcal{E}(\boldsymbol{\hbar}_A)$ and $\mathcal{E}(\boldsymbol{\hbar}_B)$ nearly aligned when $\kappa_{1,2}$ are sufficiently large. (Comparing Fig. 3a,b, $\mathcal{C}$ of system-I drops faster when $v_1$ deviates from $v_e$. This is attributed to the profiles of the extended states, which make them more sensitive to $v_1$.) In summary, with a sufficiently large value of $\kappa_{1,2}$, the characteristics of the OBC eigenstates are still dominated by the properties of ED even when $\eta(\boldsymbol{\hbar}_A)$ and $\eta(\boldsymbol{\hbar}_B)$ are slightly different.

To corroborate, we compute for each state a quantity

$$\mathcal{G} := \frac{1}{2N}\frac{\sum_{n=1}^{2N} n|\phi_n|}{\sum_{n=1}^{2N}|\phi_n|} \tag{4}$$

with $\phi_n$ being the $n$-th entry of an OBC eigenstate, and 2*N* representing the total site number. $\mathcal{G}$ effectively gauges the spatial distribution of an OBC eigenstate. In Fig. 3c (e), we see that $\mathcal{G} \cong \frac{1}{2}$ ($\mathcal{G}$ is near zero) at non-zero $\kappa_{1(2)}$, indicating all modes are extended (skin) modes at the ED. The similar results are also observed near the ED, i.e., $v_1 = -0.765$, as shown in Fig. 3d,f.

**Observation of ED-enabled response and dynamics**

We use an active mechanical system, which is capable of realizing sophisticated non-Hermitian parameters[25–30], to experimentally realize the two double-chain systems (Extended Data Fig. 1a). The one-way hopping is realized by electronic control. The natural frequency for each oscillator is 13.09 Hz, and the hopping parameter $v_1$ is $-0.765$ Hz. Note that the experimental parameters are not precisely at the ED, but as shown in Fig. 3d,f, the extended (skin) modes dominate system-I (II) when $|\kappa_{1(2)}|$ exceeds a threshold of $10^{-1}$ ($10^{-3}$). This characteristic is highly desirable for realizing ED-related phenomena, because it could be difficult for two high-dimensional eigenspaces to have identical spectra in reality.

First, we measure the frequency responses of the two chains separately and confirm that their spectra are indeed identical, as shown in Fig. 4a. The steady-state responses of the coupled double chains are shown in Fig. 4b,c. System-I (II) is excited at chain-B on site 8 (chain-A on site 6) using sinusoidal signals at 14 different frequencies as indicated in the figure. In system-I, the delocalized responses observed in chain-A confirm the presence of extended states, whereas a weak skin-mode response is seen in chain-B (Fig. 4b). The inverse is observed in system-II, where the skin-mode response in chain-B is the dominant feature (Fig. 4c). To better compare the experimental and theoretical results, we plot the frequency-integrated responses of system-I (II), where delocalized (edge-localized) responses in chain-A (chain-B) are clearly seen (Fig. 4d,e). They also agree well with the theoretical results computed using the Green's function. (We remark that, although the eigenspace spanned by skin modes (extended modes) is defective in system-I (II) at the ED, evidence of their existence can still be seen in the steady-state responses. This is due to Green's function operator at the ED also being block-triangular, as detailed in the Supplementary Section 5. It is noteworthy that Green's function at EP can also be upper-triangular[31,32].)

The ED induces two distinct dynamic effects. We reveal them by preparing two different initial states as a localized wavepacket at the center of chain-A and B, respectively, then solving the time-dependent Schrödinger equation $i\frac{d}{dt}|\psi(t)\rangle = \boldsymbol{\hbar}_{\mathrm{I,\,II}}|\psi(t)\rangle$. The result in Fig. 5a is the dissipationless dynamic responses of system-I. Here, the initial state is prepared as a single-site excitation at the center of chain-A. The wavepacket is seen to propagate symmetrically in two opposite directions and is reflected by the boundaries, which is not different from propagation in Hermitian systems[29]. Figure 5b plots the measured dynamics of system-I in our mechanical lattice. The symmetric propagation is clearly seen even in the presence of dissipation. The measured result aligns well with the theoretical one (with dissipation taken into account) shown in Fig. 5c.

When the initial state is at chain-B, the wavepacket starts to propagate in both directions, but the leftward propagation is amplified, similar to NHSE. Unlike conventional skin modes that remain localized at the boundary, this wave is reflected by the left boundary and counter-propagates across the bulk, as shown in Fig. 5d. Eventually, the wave field is propagative in the bulk and amplified to three orders of magnitude of the initial state. So we call this skin-effect amplified propagation. The fascinating dynamics of skin-effect amplified propagation are caused by the synergy of NHSE and propagation, which only occur in the proximity of the ED (Supplementary Section 6). In system-I, $\mathcal{E}(\boldsymbol{\hbar}_B)$, which is spanned by skin modes, is defective. So the system's long-time behavior is dominated by propagative modes belonging to the non-defective $\mathcal{E}(\boldsymbol{\hbar}_A)$. But an injection on chain-B can still trigger transient skin-mode responses before the waves reach the boundary, which amplify the wave toward the left. This is similar to the excitation of a single missing dimension at an EP[3]. But this amplified wave refuses to congregate at the boundary because it can escape via the propagative channels provided by the non-defective $\mathcal{E}(\boldsymbol{h}_A)$. In comparison, when the injection is on chain-A, the defective $\mathcal{E}(\boldsymbol{h}_B)$ is hidden.

The observed skin-effect amplified propagation is shown in Fig. 5e, wherein the wave propagates in both directions with the left-going wave experiencing an amplification and then reflected by the left edge. Due to system loss, the wave dissipates before reaching the right edge. The measured result sufficiently shows the key features of skin-effect amplified propagation (Supplementary Video 1). It also conforms with the theoretical result (Fig. 5f).

The second dynamic effect, propagation-enhanced skin effect, appears in system-II. When the wavepacket is injected at chain-A, it propagates both ways at first with an amplification to the leftward wave. The wave reflected by the right end (left-going after reflection) experiences considerable amplification upon reaching the left end (Fig. 5g). Clearly, the propagation-enhanced skin effect is due to a different type of synergy of NHSE and propagation near the ED, with NHSE dominating the long-time dynamics. These hallmarks are also experimentally observed, as shown in Fig. 5h, where the wave propagates in both directions in the beginning, but the branch reflected by the right boundary further grows in amplitude and eventually clings to the left edge (Supplementary Video 2). The results agree well with the theoretical calculations (Fig. 5i). In comparison, when the wavepacket is injected at the center of chain-B, the overall dynamics are akin to conventional NHSE (Fig. 5j-l).

**ED as a skin-mode controlling method**

Now recall that one necessary condition for ED is $\eta(\boldsymbol{h}_A) = \eta(\boldsymbol{h}_B)$. This condition can be broken by, e.g., offsetting $\eta(\boldsymbol{h}_B)$ by a constant $\zeta_0$ (Fig. 6a,d). The new spectra can be expressed as $\eta\left(\boldsymbol{h}_{\mathrm{I,II}}\right) = [\eta(\boldsymbol{h}_A) \cap \eta(\boldsymbol{h}_B + \zeta_0 \mathbf{I}_N)] \cup \Delta_A \cup \Delta_B$, where $\Delta_{A,B}$ are the non-overlapping parts of the spectra that belong to $\eta(\boldsymbol{h}_{A,B})$, respectively. (In finite-sized systems, the overlapping part of the spectra does not exactly overlap (inset of Fig. 6a). But the differences are negligible when the lattice is sufficiently large such that the OBC spectra approach continuum.) Apparently, the ED is now the alignment of certain portions of the two eigenspaces and does not involve the states with eigenvalues belonging to $\Delta_{A,B}$. This is confirmed in Fig. 6b,c, in which $\Delta_{A(B)}$ are populated by extended (skin) modes, whereas the states belonging to the overlapping parts remain dominated by the ED. In other words, extended and skin modes stably coexist in massive amount ($\mathcal{O}(N)$), and their ratios are tunable by simply adjusting $\zeta_0$. This functionality uniquely emerges when the system is near ED, where the topological correspondence of NHSE can be violated. It is beyond the capability of any existing non-Hermitian systems, wherein NHSE originates from the non-trivial topology of the PBC spectrum[20,33] so it cannot be turned off without fundamentally changing the system's symmetry class[34]. Competing mechanisms, e.g., magnetic flux that triggers Landau quantization[35,36] and disorders that induce Anderson localization[37], can partially control NHSE, but they rely on the delicate balance of multiple physical effects and, hence, are far more difficult to implement and control.

The effectiveness of this scheme is validated in our experiments. The onsite natural frequency of chain-B is increased from 13.09 Hz to 14.12 Hz, causing a mismatch in the frequency ranges of the eigenvalues of chain-A and chain-B. Figure 6e presents the steady-state response for system-I, where the excitation is a mono-frequency signal at 11.1 Hz, which belongs to $\Delta_A$. The response is clearly due to extended states. Shifting the excitation frequency to 15 Hz, which falls in the overlapping part of the spectrum, the response is dominated by the skin-effect amplified propagation (Fig. 6f). For an excitation at 16 Hz, which is in $\Delta_B$, the response indicates conventional NHSE (Fig. 6g). In system-II, when the excitation is at 11.1 Hz and 16 Hz, the results are due to extended and skin modes, respectively (Fig. 6h, j). For an excitation at 15 Hz, the response indicates propagation-enhanced skin effect (Fig. 6i).

**Discussion**

The ED is a condition at which the Hilbert space of a $2N \times 2N$ square matrix is $N$-fold defective, accompanied by all eigenvalues being two-fold degenerate. Such a matrix is in a Jordan canonical form composed of $N$ different Jordan cells of size 2. These conditions are distinct from conventional order-$N$ EPs, where all $N$ eigenvectors align, and the eigenvalues are $N$-fold degenerate. As a multi-dimensional generalization of EPs, ED transcends previous understandings and expectations of non-Hermitian physics and sets the stage for diverse new research directions. ED is a generic condition hinging on the block-triangular form of the Hamiltonian and is characterized by the $O(N)$ defectiveness of Hilbert space. There are no restrictions on the specifics of the Hilbert space (such as orthogonality and dimensionality) and the forms of the spectra – they can be continuous, discrete (formed by states with isolated energy), or any combination of them. ED and related phenomena are also closely related to the exceptional spectrum considered in ferromagnetic systems[38].

One surprising property of the ED formed by continua is the robustness against certain types of local perturbations, such as randomness in onsite energy or inter-chain coupling (Supplementary Section 1). Yet, in the meantime, ED is critically sensitive to specific perturbations that break the mathematical structure of the Hamiltonian, which is ideal for sensing applications. A design and the numerical results of an ED-based ultra-sensitive sensor with a broad operational bandwidth and large dynamic range are shown in Extended Data Fig. 2. More discussions on ED criticality are shown in Supplementary Section 1.

ED is realizable using different metamaterials or even with natural materials by simply stacking them in layers with purely one-way coupling, which is achievable by utilizing spin-orbit coupling in microwaves[39], electrical coupling in acoustics[40], and electrical components[41,42].

We believe the skin-propagative dynamics reported here are merely the tip of an iceberg for what phenomena ED can bring. Perhaps the most anticipated advancement is the broadband exceptional physics and functionalities – a viable solution to the narrowband characteristics inherent to all EPs. Indeed, the ED involving the coincidence of spectra without bandwidth limitation is particularly desirable for sensing[43], modal control[44], and lasing[45], which are famed applications hinging on the properties of EPs.

Future works will explore ED appearing in Hamiltonians with more general forms beyond the Jordan cell structure. Considering the diverse topological properties of EPs[46–52], ED may also be associated with exotic high-dimensional topology that governs their evolutions in parameter space and possible merger. It may also bring unexpected twists to non-Hermitian correlated systems, wherein EPs are predicted to produce negative entanglement entropy and exceptional bound states [53].

**Acknowledgments**

This work was supported by the National Key R&D Program (2022YFA1404400), the National Natural Science Foundation of China (T2525002), the Hong Kong Research Grants Council (RFS2223-2S01, 12301822, 12300925, JRFS2526-2S07), and the Hong Kong Baptist University (RC-RSRG/23-24/SCI/01, RC-SFCRG/23-24/R2/SCI/12). K.S. and M.S. were supported by JST CREST (JPMJCR19T2). M.S. was supported by JSPS KAKANHI (JP24K00569). K.S. was supported by JST SPRING (JPMJSP2110) and JSPS KAKENHI Grant (JP25KJ1632). Z.Y. was also sponsored by the National Key R&D Program (2023YFA1407500) National Natural Science Foundation of China (12322405, 12104450). Zhen Li is grateful to Dr. Zhao Feng for his valuable suggestions on figure organization. K.S., M.S., and G.M. thank the hospitality of the Simons Center for Geometry and Physics at Stony Brook University.

**Author contributions**

G.M. conceived and supervised the research; Z.L., K.S., C.L., Z.Y., M.S., and G.M. performed theoretical analysis; Z.L., R.C., and X.W. performed the experiments; all authors analyzed and discussed the results and contributed to the manuscript.

**Competing interests**

The authors declare no competing interests.

**Fig. 1 | ED as a generalization of EP. a**,**b**, When a two-level system approaches an EP (marked by the magenta color), the eigenvalues gradually become identical (**a**) and the eigenvectors coalesce (**b**). **c**, Conceptual depictions of the coalescence of two eigenspace when the diagonal entries $h_A$ and $h_B$ in the triangular matrix are replaced by the Hermitian matrices $\mathbf{h}_A$ and $\mathbf{h}_B$, and the hopping changes from the number $\kappa$ to the matrix $\boldsymbol{\kappa}$ accordingly. The red and blue hyper-cubes represent the orthogonal eigenspace of $\mathbf{h}_A$ ($\mathcal{E}(\mathbf{h}_A)$) and $\mathbf{h}_B$ ($\mathcal{E}(\mathbf{h}_B)$). The magenta hyper-cube represents the coalesced eigenspace of equation (1) ($\mathcal{E}(\mathbf{H})$). **d**, Similar to **c**, but $\mathbf{h}_B$ changes to the non-Hermitian matrix. The skewed eigenspace of $\mathbf{h}_B$ is represented as the blue hyper-parallelepipeds. This results in a highly defective but orthogonal Hilbert space. **e**, Similar to **d**, but the hopping direction is reversed, inducing the Hermitian eigenspace coalescing with a skewed non-Hermitian one.

**Fig. 2 | ED-induced anomalous NHSE in one-dimensional double-chain lattices. a**, The two double-chain models with block-triangular Hamiltonians (equation (2)). The two chains are one-way coupled by $\kappa_1$ or $\kappa_2$. The lattice site indices are illustrated for the unit cell enclosed by the green box. **b**, The PBC spectra of system-I and II, which are the union of the spectra of chain-A ($\eta(\boldsymbol{\mathcal{H}}_A)$) and chain-B ($\eta(\boldsymbol{\mathcal{H}}_B)$). **c**, The identical generalized Brillouin zone of two systems, which consists of two sub-generalized Brillouin zones $\beta_A$ and $\beta_B$. **d**, The OBC spectra of two systems, which are the union of the spectra of chain-A ($\eta(\boldsymbol{\hbar}_A)$) and chain-B ($\eta(\boldsymbol{\hbar}_B)$) and pairwise degenerate everywhere, as shown in the inset. The PBC spectra, OBC spectra, and generalized Brillouin zone of system-I and II are all identical. **e**,**f**, The OBC eigenstates of system-I (**e**) and system-II (**f**), ordered in ascending order of the eigenvalues. The OBC bandgap and zero-energy edge modes are irrelevant to our study and are omitted for clarity. Parameters in calculations are $w = -1.565$, $v_2 = -0.885$, $\delta = -0.453$, $\kappa_1 = \kappa_2 = -1.081$, $v_1 = v_e = -\sqrt{(v_2 - \delta)(v_2 + \delta)}$. ($v_e$ is the parameter at which the two systems are at ED, which is an irrational number due to the square root. We kept 30 digits in the calculations.)

**Fig. 3 | The properties of the Hilbert space and eigenstates at or near the ED. a**,**b**, Cosine similarities $\mathcal{C}$ between the two coalescing eigenspaces in system-I (**a**) and system-II (**b**). The insets of **a** and **b** plot $\mathcal{C}$ at the ED ($v_1 = v_e$, black dots) and at the experimental parameter

($v_1 = -0.765$, red dots). The two eigenspaces can be nearly aligned ($\mathcal{C} \cong 1$) when the parameters slightly deviate from the ED condition, provided $\kappa_{1,2}$ is sufficiently large. **c**,**d**, Spatial characteristics of continuum eigenstates at ED (**c**) and slightly away from ED (**d**) for system-I. In **d**, extended states dominate with sufficiently large $\kappa_1$, despite the deviation from the ED. **e**,**f**, Similar to **c**,**d**, but for system-II. Here, skin modes dominate near the ED. The quantity $\mathcal{G}$ is defined in equation (4).

**Fig. 4 | The steady-state responses at the ED. a**, Frequency responses of chain-A and chain-B in isolation show that they have an identical spectrum. The upper and lower panels are the computed (with Green's function, normalized) and measured results, respectively. **b**,**c**, Experimentally measured steady-state responses of system-I (**b**) and system-II (**c**) excited by sinusoidal signals at 14 different frequencies. The shaded red and blue lines represent the amplitude of chain-A and chain-B, respectively. The source position is fixed at site 8 (site 6) in the measurement of system-I (II), as indicated by the blue dots. **d**,**e**, The steady-state responses summed over all measured frequencies of system-I (**d**) and system-II (**e**). The frequency-integrated responses at each site is defined as $A_n = \sum_{i=1}^{14} |\theta_n(f_i)|$, where *n* represents the site index, and *i* indices the measured frequency *f*. The dashed lines and solid lines with shading plot computed and measured results, respectively. In the experiment, the hopping is $v_1 = -0.765$ Hz, which slightly deviates from $v_e$. The onsite natural frequency is 13.09 Hz. All other parameters are the same as those in Fig. 2.

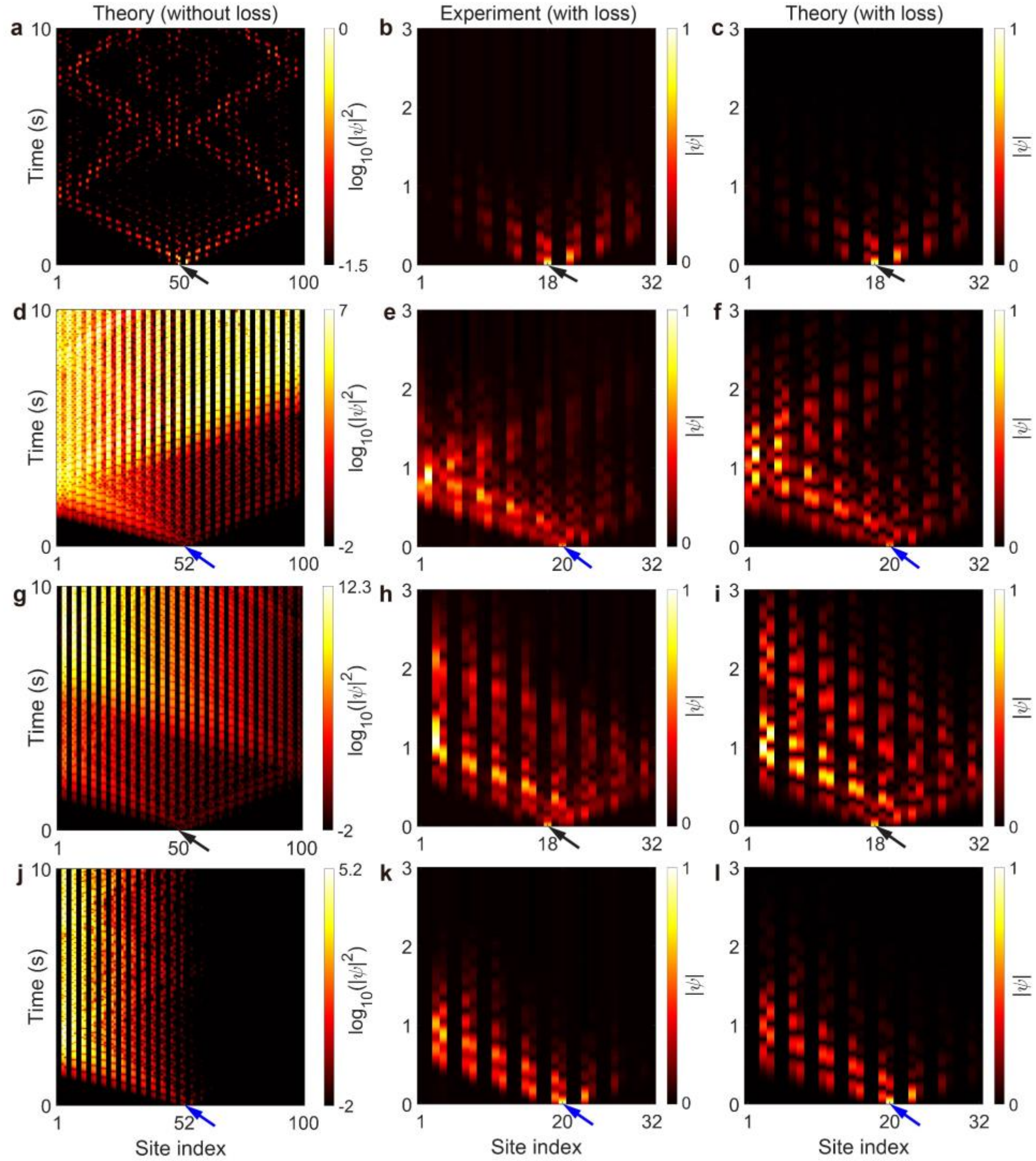

**Fig. 5 | ED-induced skin-propagative synergized dynamics. a**, The dissipationless time evolution of the probability density in system-I when a wavepacket is injected is at the center of chain-A (the source position is marked by the black arrow). **b**,**c**, The corresponding results of experimental measurement (**b**) and theoretical calculation using experimental parameters (**c**). Symmetric propagation is clearly observed, indicating the appearance of extended modes. Owing to dissipation, the waves vanish in 1 s. **d**-**f**, The skin-effect amplified propagation in system-I when the source is at the center of chain-B (marked by the blue arrow). **g**-**i**, The propagation-enhanced skin effect in system-II when the source is at the center of chain-A. **j**-**l**, The typical non-Hermitian skin dynamics in system-II when the source is at the center of chain-B. The parameters are the same as those used in Fig. 4. In the theoretical results in (**a**, **d**, **g**, **j**), the system size is 100, the time evolutions are dissipationless, and the colormaps are assigned to the logarithm (base 10) of the probability density. In (**c**, **f**, **i**, **l**), the onsite term is 13.09-0.34i to match the experimental condition, and the colormaps are in linearscale.

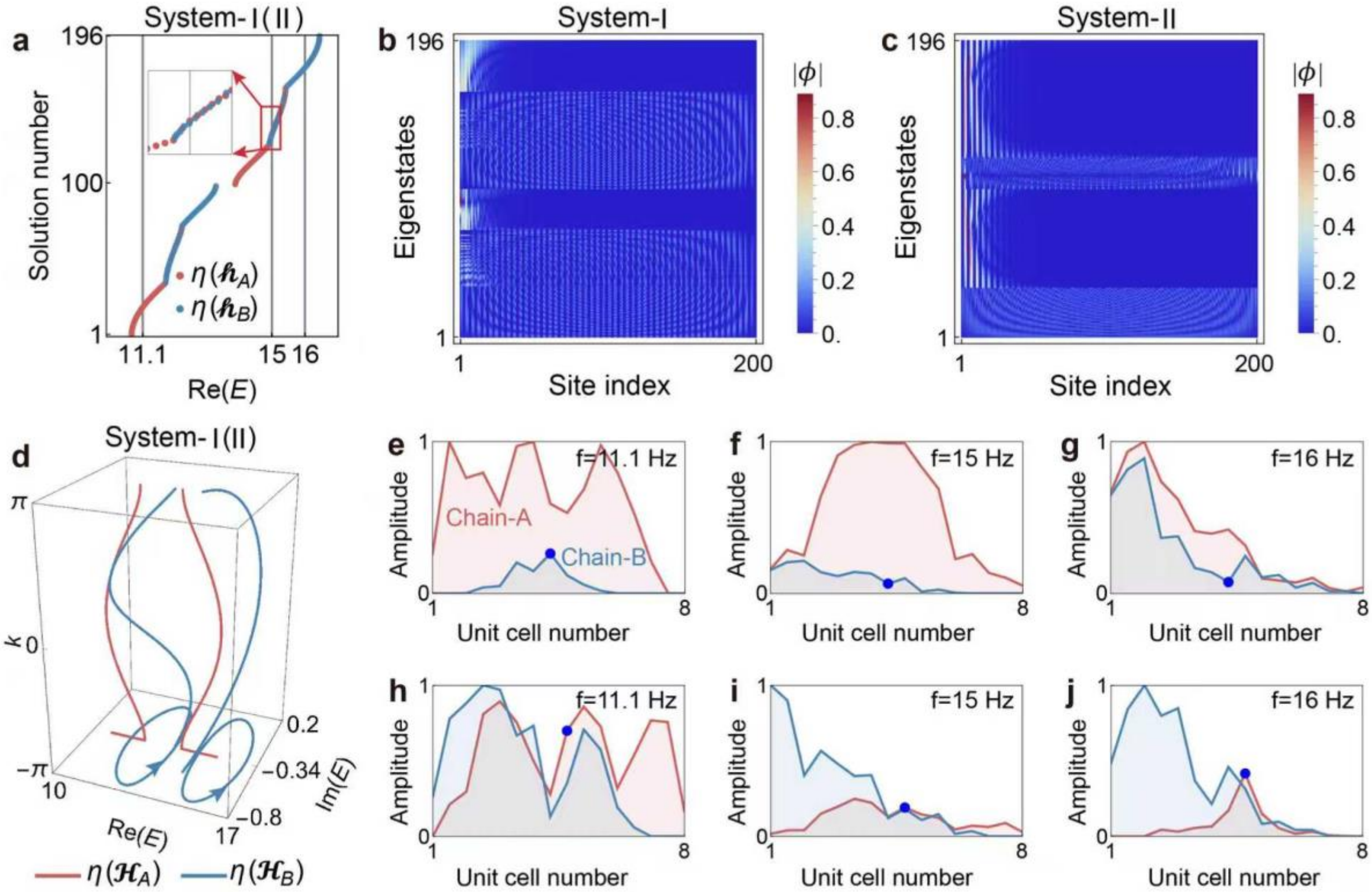

**Fig. 6 | Controlling extended and skin modes using ED. a**,**d**, The OBC and PBC spectra of system-I and II with an offset in the spectrum of chain-B. **b**,**c**, The eigenstates of the continuous bands of system-I (**b**) and II (**c**). The eigenstates are ordered in ascending order of the OBC eigenvalues. Extended modes and skin modes coexist in massive amounts. Their numbers are arbitrarily tunable by the offset. **e**-**g**, Experimentally measured steady-state response of system-I at 11.1 Hz (**e**), 15 Hz (**f**), 16 Hz (**g**). **h**-**j**, The experimental results for system-II. In **e-j**, the red and blue lines with shading plot the steady-state responses of chain-A and B, respectively. The excitation position is marked by blue dots. The natural frequencies of chain-A and chain-B are $f_A = 13.09$ Hz, $f_B = 14.12$ Hz. All other parameters are the same as those in Fig. 4.

## Methods

### Eigenstates of the block-triangular matrices

Here, we take the system-I as an example and demonstrate the exact form of eigenstates and their relationships with the eigenvalues. The OBC Hamiltonian of system-I is $\boldsymbol{h}_{\mathrm{I}} = \begin{pmatrix} \boldsymbol{h}_A & \boldsymbol{\kappa}_1 \\ 0 & \boldsymbol{h}_B \end{pmatrix}$, where $\boldsymbol{h}_A$ and $\boldsymbol{h}_B$ represent the OBC Hamiltonian of chain-A and chain-B, respectively (Note that the specific form of $\boldsymbol{h}_{\mathrm{I}}$ here requires re-numbering the sites, which simply amounts to a similarity transform and has no effect on the system). $\boldsymbol{h}_{\mathrm{I}}$ is also block-triangular and both $\boldsymbol{h}_A$ and $\boldsymbol{h}_B$ are diagonalizable, so there exists an invariant subspace[22]. Although the hopping between the two chains is a diagonal matrix in this example, we demonstrated that the exact form of the hopping matrix plays no role in the formation of the ED (Supplementary Section 1). The eigenproblem gives three equations

$$\begin{pmatrix} \boldsymbol{h}_A & \boldsymbol{\kappa}_1 \\ \mathbf{0} & \boldsymbol{h}_B \end{pmatrix} \begin{pmatrix} |u_A\rangle \\ |u_B\rangle \end{pmatrix} = E_{\mathrm{I}} \begin{pmatrix} |u_A\rangle \\ |u_B\rangle \end{pmatrix}, \tag{5}$$

$$\boldsymbol{h}_A |u_A\rangle + \boldsymbol{\kappa}_1 |u_B\rangle = E_{\mathrm{I}} |u_A\rangle, \tag{6}$$

$$\boldsymbol{h}_B |u_B\rangle = E_{\mathrm{I}} |u_B\rangle. \tag{7}$$

Theorem of invariant subspace suggests that the spectrum of $\boldsymbol{h}_{\mathrm{I}}$ is the union of the spectrum of $\boldsymbol{h}_A$ and $\boldsymbol{h}_B$, $\eta(\boldsymbol{h}_{\mathrm{I}}) = \eta(\boldsymbol{h}_A) \cup \eta(\boldsymbol{h}_B)$. Then for $|\phi\rangle = \begin{pmatrix} |u_A\rangle \\ |u_B\rangle \end{pmatrix}$, where $|u_A\rangle$ and $|u_B\rangle$ have the same dimension, to be right eigenvectors of $\boldsymbol{h}_{\mathrm{I}}$, there are three possibilities for the eigenvalue $E_{\mathrm{I}}$:

When $E_{\mathrm{I}} \in \eta(\boldsymbol{h}_A)$ and $E_{\mathrm{I}} \notin \eta(\boldsymbol{h}_B)$, we have $|u_B\rangle = 0$ from equation (7). Thus, equation (6) becomes $\boldsymbol{h}_A |u_A\rangle = E_{\mathrm{I}} |u_A\rangle$. Therefore, $|u_A\rangle$ is either a zero vector (a trivial solution) or a right eigenvector of $\boldsymbol{h}_A$. So such eigenvectors become $|\phi\rangle = \begin{pmatrix} |u_A\rangle \\ 0 \end{pmatrix}$.

When $E_{\mathrm{I}} \notin \eta(\boldsymbol{h}_A)$ and $E_{\mathrm{I}} \in \eta(\boldsymbol{h}_B)$, $(\boldsymbol{h}_A - E_{\mathrm{I}})^{-1}$ exists, so equation (6) implies $|u_A\rangle = -(\boldsymbol{h}_A - E_{\mathrm{I}})^{-1} \boldsymbol{\kappa}_1 |u_B\rangle$. Thus, if $|u_B\rangle = 0$, then we also have $|u_A\rangle = 0$, which leads to a trivial solution. So, to obtain a non-trivial solution, $|u_B\rangle$ must be a non-zero vector. Then, equation (7) suggests that $|u_B\rangle$ is the eigenstates of $\boldsymbol{h}_B$, so the eigenvectors are $|\phi\rangle = \begin{pmatrix} -(\boldsymbol{h}_A - E_{\mathrm{I}})^{-1} \boldsymbol{\kappa}_1 |u_B\rangle \\ |u_B\rangle \end{pmatrix}$. Those two cases are shown in Fig. 6a,b.

The third possibility is $\eta(\boldsymbol{h}_A) = \eta(\boldsymbol{h}_B)$. So for $E_{\mathrm{I}} \in \eta(\boldsymbol{h}_A)$, there is also $E_{\mathrm{I}} \in \eta(\boldsymbol{h}_B)$. If $|u_B\rangle \neq 0$, then $|u_B\rangle$ is the right eigenstate of $\boldsymbol{h}_B$ from equation (7). Since there is also $E_{\mathrm{I}} \in \eta(\boldsymbol{h}_A)$, there exists a left eigenstate $\langle v_A|$ of $\boldsymbol{h}_A$ satisfying $\langle v_A|\boldsymbol{h}_A = E_{\mathrm{I}}\langle v_A|$. Thus, equation (6) implies

$$\langle v_A|\boldsymbol{\kappa}_1|u_B\rangle = 0. \tag{8}$$

Since equation (8) is not generally held, so we must require $|u_B\rangle = 0$. (In certain cases, equation (8) holds. But these are non-generic cases that require $\boldsymbol{\kappa}_1$ or $|u_B\rangle$ to have special structures.) Then, from equation (6), $|u_A\rangle$ is a right eigenstate of $\boldsymbol{h}_A$, and $|\phi\rangle = \begin{pmatrix} |u_A\rangle \\ 0 \end{pmatrix}$. This is exactly the situation of the exceptional deficiency (ED), as shown in Fig. 2d,e.

We have shown that equation (8) is a necessary condition for the absence of ED so far, the converse can also be proved: for any $E_{\mathrm{I}} \in \eta(\boldsymbol{h}_A) = \eta(\boldsymbol{h}_B)$, ED does not occur if equation (8) holds for all the right eigenvectors $|u_B, b\rangle$ of $\boldsymbol{\hbar}_B$ and all the left eigenvectors $\langle v_A, a|$ of $\boldsymbol{\hbar}_A$ associated

$$\langle v_A, E_\mathrm{I}, a|\boldsymbol{\kappa}_1|u_B, E_\mathrm{I}, b\rangle = 0, \text{ for } \forall a, b, E_\mathrm{I}, \tag{9}$$

where $a$, $b$ index the eigenvectors associated with $\mathcal{E}(\boldsymbol{\hbar}_A)$ and $\mathcal{E}(\boldsymbol{\hbar}_B)$. Indeed, under equation (9), we can construct all the linearly-independent right eigenvectors of $\boldsymbol{\hbar}_\mathrm{I}$ with each eigenvalue $E_\mathrm{I}$ as

$$|\phi\rangle = \begin{pmatrix} |u_A, E_\mathrm{I}, a\rangle \\ 0 \end{pmatrix}, \begin{pmatrix} -\sum_{E \neq E_\mathrm{I}} (E - E_\mathrm{I})^{-1} \sum_a \langle v_A, E, a|\boldsymbol{\kappa}_1|u_B, E_\mathrm{I}, b\rangle \, |u_A, E, a\rangle \\ |u_B, E_\mathrm{I}, b\rangle \end{pmatrix}, \tag{10}$$

which span the total Hilbert space. Here,

$$\boldsymbol{\hbar}_A = \sum_{E \in \eta(\boldsymbol{\hbar}_A)} \sum_a E |u_A, E, a\rangle\langle v_A, E, a|, \tag{11}$$

$$\boldsymbol{\hbar}_B = \sum_{E \in \eta(\boldsymbol{\hbar}_B)} \sum_b E |u_B, E, b\rangle\langle v_B, E, b|, \tag{12}$$

$$\langle v_A, E, a|u_A, E', a'\rangle = \delta_{E,E'}\delta_{a,a'}, \langle v_B, E, b|u_B, E', b'\rangle = \delta_{E,E'}\delta_{b,b'}. \tag{13}$$

Hence, equation (8) is a necessary and sufficient condition for the absence of the ED; equivalently, the negation of equation (8) is that for the presence of the ED.

**Cosine similarity**

Cosine similarity evaluates the similarity between two linear spaces. In our work, the two eigenspaces, denoted $\mathcal{E}(\boldsymbol{\hbar}_A)$ and $\mathcal{E}(\boldsymbol{\hbar}_B)$, are spanned by the eigenvectors corresponding to $\eta(\boldsymbol{\hbar}_A)$ and $\eta(\boldsymbol{\hbar}_B)$, respectively (Each column of $\mathcal{E}\big(\boldsymbol{\hbar}_{A(B)}\big)$ represents a right eigenvector $\Big|\phi_{m_{A(B)}}\Big\rangle$). The cosine similarity is defined as $\mathcal{C}\big(\mathcal{E}(\boldsymbol{\hbar}_A), \mathcal{E}(\boldsymbol{\hbar}_B)\big) = \frac{\langle \mathcal{E}(\boldsymbol{\hbar}_A), \mathcal{E}(\boldsymbol{\hbar}_B)\rangle_F}{\|\mathcal{E}(\boldsymbol{\hbar}_A)\|_F \|\mathcal{E}(\boldsymbol{\hbar}_B)\|_F}$, where $\langle \mathcal{E}(\boldsymbol{\hbar}_A),\ \mathcal{E}(\boldsymbol{\hbar}_B)\rangle_F = \sum_{m_{A(B)}=1}^{M_{A(B)}} \big|\langle \phi_{m_A}|\phi_{m_B}\rangle\big|$ is the Frobenius inner product (Here, $M_A + M_B$ gives the total number of bulk states, and in our systems $M_A = M_B = (2N - 4)/2$, $2N$ is the lattice size, and there are $(2N - 4)$ modes in the continuum bands, 4 modes are in-gap edge modes), $\langle \phi_{m_A}| = \big(|\phi_{m_A}\rangle\big)^\dagger$, $\big\| \mathcal{E}\big(\boldsymbol{\hbar}_{A(B)}\big)\big\|_F = \sqrt{\sum_{m_{A(B)}}^{M_{A(B)}} \sum_{n=1}^{2N} \Big|\phi_{m_{A(B)}n}\Big|^2}$ is the Frobenius norm of $\mathcal{E}\big(\boldsymbol{\hbar}_{A(B)}\big)$. We do not differentiate parallel and anti-parallel spaces, with the resulting similarity ranging from 0 to 1, indicating an orthogonal or parallel relation between the two eigenspaces.

**Experimental setup**

The experimental system is based on the active mechanical lattice, a proven technology for realizing non-Hermitian lattices[25–29]. As shown in Extended Data Fig. 1a, the lattice has eight unit cells, each with four sites, which are realized by harmonic oscillators with a single rotational mode. A selected unit cell is highlighted by the white box, and its schematic diagram (of system-II) is depicted in Extended Data Fig. 1b.

The setup of a single oscillator is shown in Extended Data Fig. 1c. Each oscillator is comprised of a rigid arm loaded with weights, anchored with two springs (length: 8.0 cm, stiffness coefficient: 50.9 N/m), and fixed on a programmable motor. The excitation signal is sent to the motor through the microcontroller board. Then, the real-time rotation angle of the motor is detected by the magnetic sensor.

The oscillators are properly connected by springs with designated spring constants to realize reciprocal hopping. Asymmetric hopping is achieved through a programmed feedback control that drives specific motors. The hopping parameters and the onsite orbital are calibrated at a unit cell, as shown in Extended Data Fig. 1b. Take the calibration of non-reciprocal hopping in chain-B as an example, a chirp signal is sent to motor 4. Thereafter, a microcontroller receives the real-time rotation angle $\theta_4$ of motor 4, and then it drives motor 3 with an additional torque $\tau_3 = a_{34}\theta_4$ besides the reciprocal hopping provided by the tensioned springs, where $a_{34}$ is a controlling parameter. This effectively emulates an additional linear spring for oscillator 3. The non-reciprocal hopping can be retrieved by fitting the measured frequency response using Green's function, which is described in detail in refs.[25,27,29]. Sites 1 and 2 remain still under this condition because they are unidirectionally coupled to sites 3 and 4, respectively. In the entire lattice, the two chains are coupled through purely one-way hopping; they are coupled solely through such programmed feedback with no physical spring connecting them. For steady-state response measurements, the source was activated for 20 s, and the data sampling rate was 500 Hz.

## 1. Robustness and fragility of the ED

The exceptional deficiency (ED) shows rich behaviors under different kinds of perturbation and variations of parameters. Here, we present a comprehensive discussion.

### 1.1. Robustness

One would be tempted to think that ED being such a unique and unorthodox situation, any perturbation would have devastating effects. Surprisingly, this is not necessarily the case. Because ED-induced phenomenon affects a spectral continuum, it is quite robust against local perturbations. For example, Fig. S1(a, b) presents the change of the cosine similarity when onsite random disorders are introduced to different percentages of sites in system-I. Therein, $\alpha$ denotes the magnitude of randomness. It is seen that the cosine similarities remain at a level close to unity even in the presence of substantial disorder. The spectral mismatches at the band edges are a major contributing factor to the reduction in the cosine similarities. This result is corroborated by the OBC eigenstates that remain stably extended modes, as shown in Fig. S1(c, d). Figure S1(e-h) shows the corresponding results for system-II.

In the OBC systems, when the density of states is sufficiently large such that the bands become semi-continuum, and the presence of disorder merely shuffles the order of eigenstates, the spectra of the two eigenspaces still overlap. And because the formation of ED does not care about the specifics of the eigenstates, the ED-induced phenomena survive such disorder.

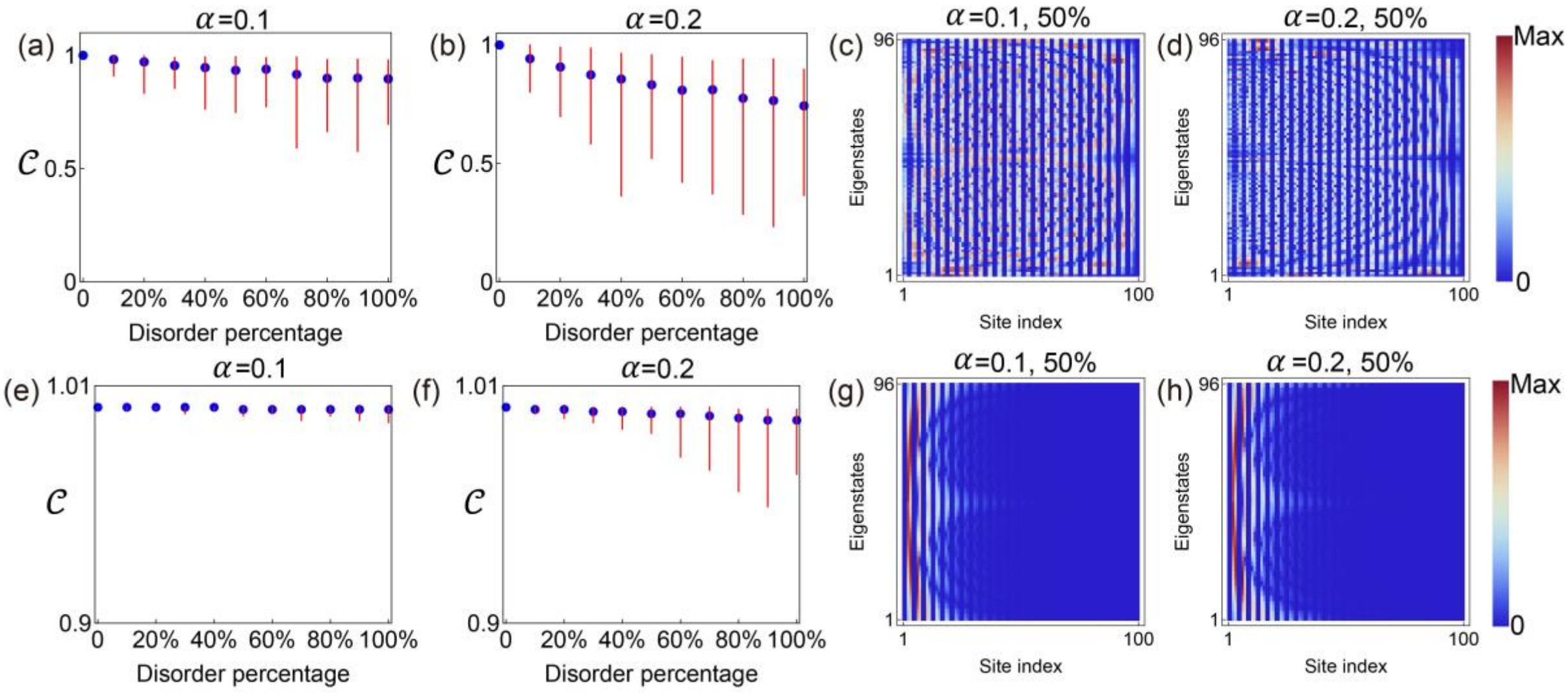

**Fig. S1 | Robustness of the ED.** (a) and (b) The cosine similarities between eigenspaces in system-I when different percentages of disorders, with amplitude $\alpha = 0.1, 0.2$ are introduced to the onsite energies. (c) and (d) Two selected sets of OBC eigenstates under disorder perturbations. (e) to (h) The corresponding results in system-II. In (a, b) and (e, f), the blue points indicate the cosine similarities averaged over 100 configurations, and the red bars represent the range of all data sets. The onsite term is 13.09-0.34i, and other parameters are the same as Fig. 2 in the main text.

In addition, although the hopping matrices in system-I and II of the main text are diagonal, we demonstrated that the exact form of the hopping matrix plays no role in the formation of the

ED.

Here, we consider two additional hopping configurations. In Fig. S2(a), $\boldsymbol{\kappa}$ is a diagonal matrix with random numbers on the diagonal. In Fig. S2(e), the arrangement of the interchain hopping produces a non-diagonalizable $\boldsymbol{\kappa}$. Yet, it is seen that ED appears for both cases [Fig. S2(b, c) and (f, g)]. These results indicate that when the diagonal elements are replaced with a series of random real numbers or the hopping matrix cannot be diagonalized, the formation of ED remains largely unaffected. The only observable effect is a slight increase in the critical values at which cosine similarity transitions occur, as shown in Fig. S2(d, h).

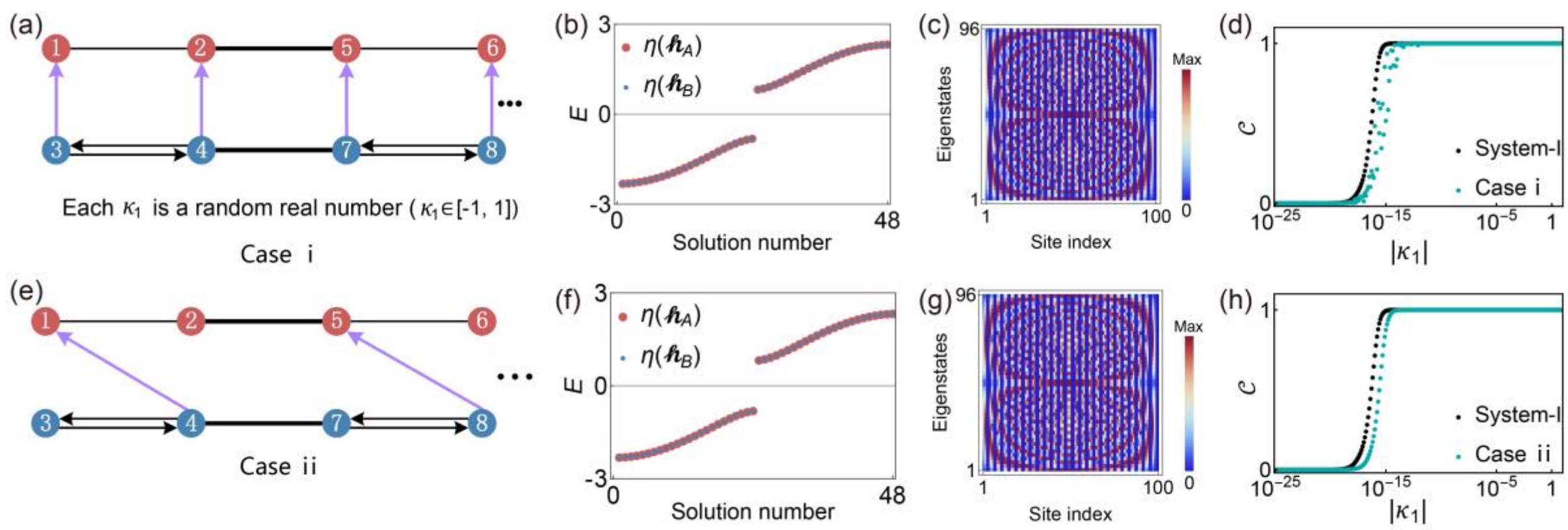

**Fig. S2 | Two additional configurations of interchain hopping.** (a) The inter-chain hoppings are real random number ($\kappa_1 \in \mathrm{rand}[-1,1]$). (b-c) The corresponding OBC energy spectrum and eigenstates show the presence of ED. (d) The variation of the cosine similarity metric as a function of $|\kappa_1|$. (e-h) The results of a non-diagonalizable $\boldsymbol{\kappa}$, where $\kappa_1$ is the same as in the main text.

### 1.2 Criticality

As mentioned in the main text, the block-triangular form of the Hamiltonian is crucial for the emergence of the ED. Consequently, the ED is fragile against perturbations that break the block-triangular form. Figure S3(a, b) shows the PBC and OBC spectra of system-I when the inter-chain hopping is no longer strictly one-way. As $\kappa_2$ increases from 0 to 0.1, the spectra $\eta(\boldsymbol{\hbar}_A)$ and $\eta(\boldsymbol{\hbar}_B)$ no longer overlap (in fact, it is no longer possible to clearly separate the two eigenspaces into Bloch waves and non-Bloch waves). And the OBC eigenvectors gradually evolve into skin modes [Fig. S3(c-f)]. In system-II, when the same perturbation is applied as $\kappa_1$, the PBC and OBC spectra remain identical to those of system-I [Fig. S3(g, h)]. The skin modes eventually become identical to the ones of system-I [Fig. S3(i-l)]. However, the differences are less pronounced because the modes of system-II are already skin modes at ED to begin with.

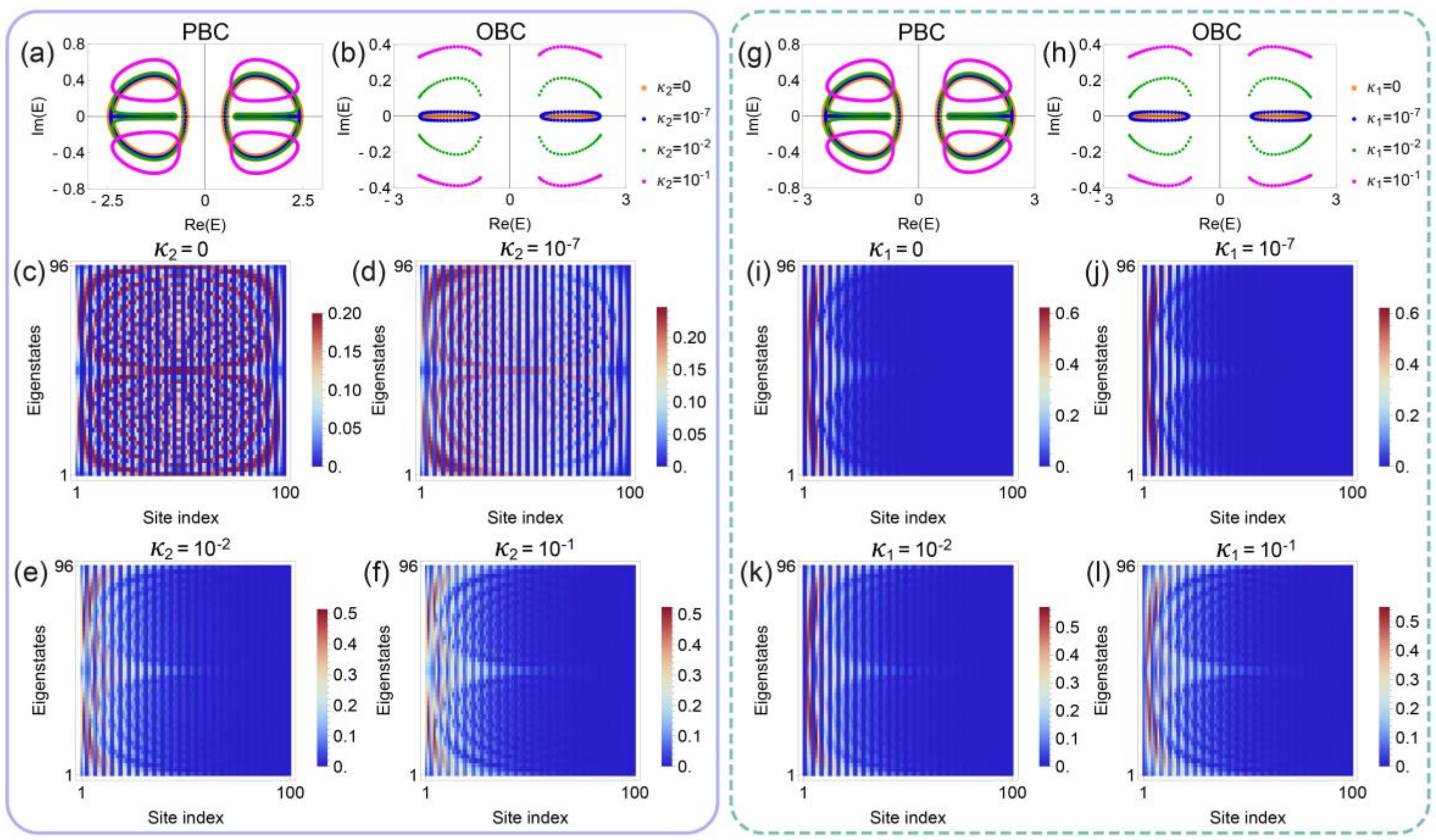

**Fig. S3 | Fragility of the ED against perturbations that break the block-triangular form of the Hamiltonian.** (a) The PBC and (b) OBC spectra of system-I when a perturbation is introduced as the A-to-B inter-chain hopping, denoted $\kappa_2$. (c-f) The corresponding OBC eigenstates. (g) PBC and (h) OBC spectra of system-II when a perturbation is introduced as the B-to-A inter-chain hopping, denoted $\kappa_1$. (i-l) The corresponding OBC eigenstates. The parameters used in the calculations are identical to Fig. 2 in the main text with zero onsite terms.

## 2. "ED curves"

Exceptional points (EPs) are known to form continuous curves under suitable conditions. Although the ED is completely different from EPs in that it is not a point in the spectrum, it can appear over a continuous region of parameters. Figure S4 illustrates the similarity as a function of system parameters $v_1$ and $\delta$ in our systems. For system-I, the ED exists when the equation $v_1^2 + \delta^2 = v_2^2$ is satisfied, which is a circle in the $v_1\delta$-space. A similar ED circle is also found in system-II.

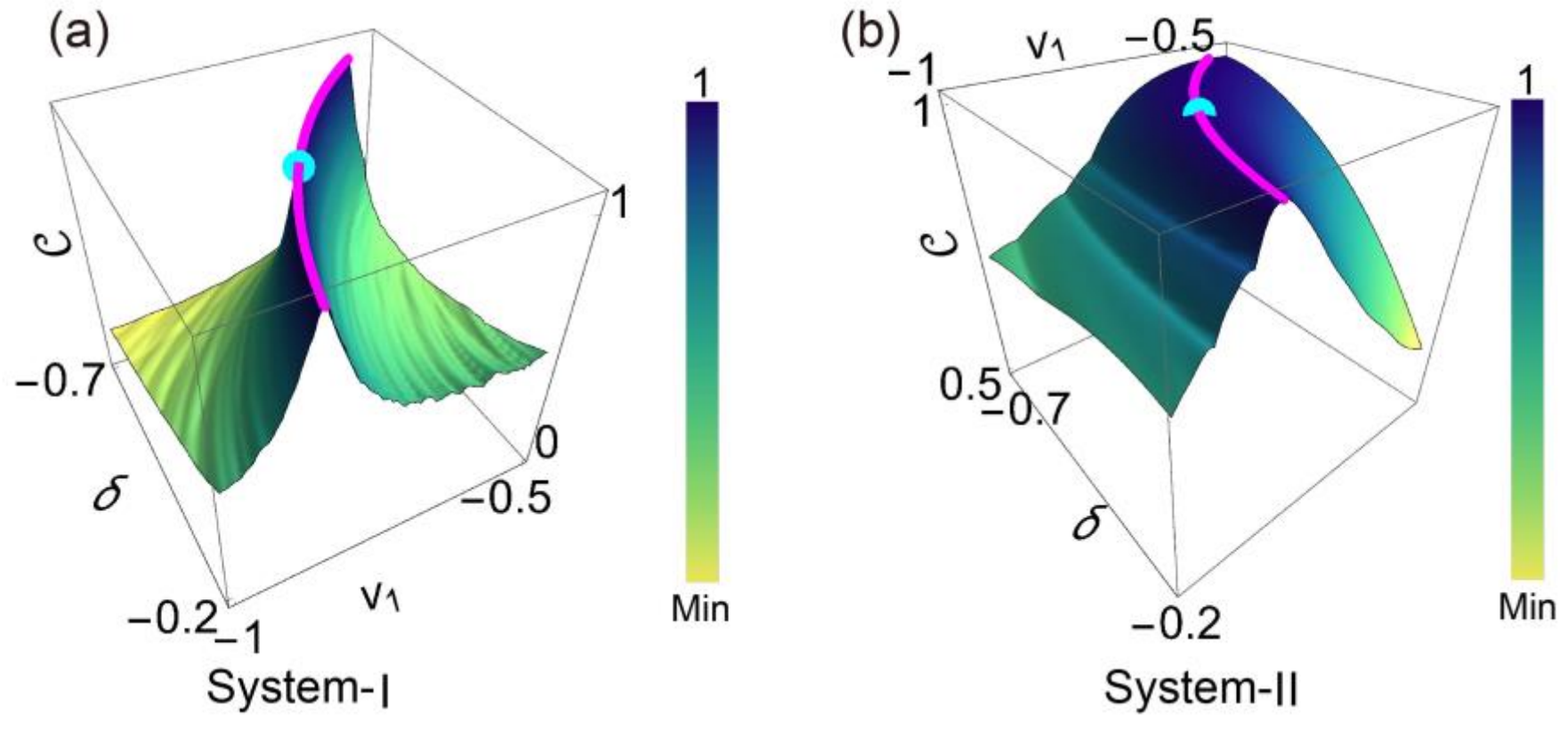

**Fig. S4 | Continuous ED curves.** The emergence of ED circles (marked by the magenta color) in system-I (a) and II (b). The equation of the ED curve is $v_1 = -\sqrt{(v_2 - \delta)(v_2 + \delta)}$. The

cyan dots represent the positions of the ED in Fig. 2 in the main text. Other parameters are the same as Fig. 2 in the main text with zero onsite terms. Only a quarter of the circle is shown.

## 3. Additional models

To demonstrate the generality of the ED, we present similar effects in several other models. Figure S5(a) depicts systems consisting of double-chain Hatano-Nelson (HN) models. The intra-chain hopping parameters in chain-a and b have the same amplitude but opposite signs, i.e., $\delta_a = -\delta_b$. When the two chains are isolated, their PBC and OBC spectra coincide (Fig. S5(b, c)). The windings of the PBC spectra are opposite (Fig. S5(b)), which indicates that the skin modes in chain-a and chain-b are localized at the left edge and right edge, respectively. When the two chains are one-way coupled, it is straightforward to see that the Hamiltonian has a block-triangular form, and the condition for ED is met. As a consequence, system-1 (2) is exclusively populated by leftward (rightward) skin modes, as shown in Fig. S5(d, e).

When $\delta_a \neq -\delta_b$, the PBC and OBC spectra of the two chains no longer coincide, as depicted in Fig. S5(f, g). The OBC eigenstates of the two systems are shown in Fig. S5(h, i). The OBC eigenstates corresponding to the overlapping spectra of system-1 and 2 are dependent on the direction of the inter-chain hopping, whereas the non-overlapping eigenstates retain the profile of leftward skin modes in system-2, which belong to chain-a. In particular, both leftward and rightward skin modes exist in system-2, and the ratio of the two kinds of skin modes is tunable by changing the spectra of chain-a and chain-b. These observations are consistent with the theoretical predictions of the ED.

To further extend the idea, we also explore a model that couples a Hermitian SSH model and a non-Hermitian HN model, denoted as system-3 and system-4 and are shown in Fig. S6(a). The PBC spectra consist of a closed loop with a clockwise winding and a real spectrum, as shown in Fig. S6(b). The OBC spectra are comprised of two spectral lines lying on the real axis with partially overlapping energies, as illustrated in Fig. S6(c). The GBZs of those two systems are also identical (Fig. S6(d)). The eigenstates corresponding to the overlapping energies are extended states in system-3 and skin modes in system-4 (Fig. S6(e, f)).

These results indicate that ED-induced effects can be observed in many different systems, as long as the relevant conditions are met.

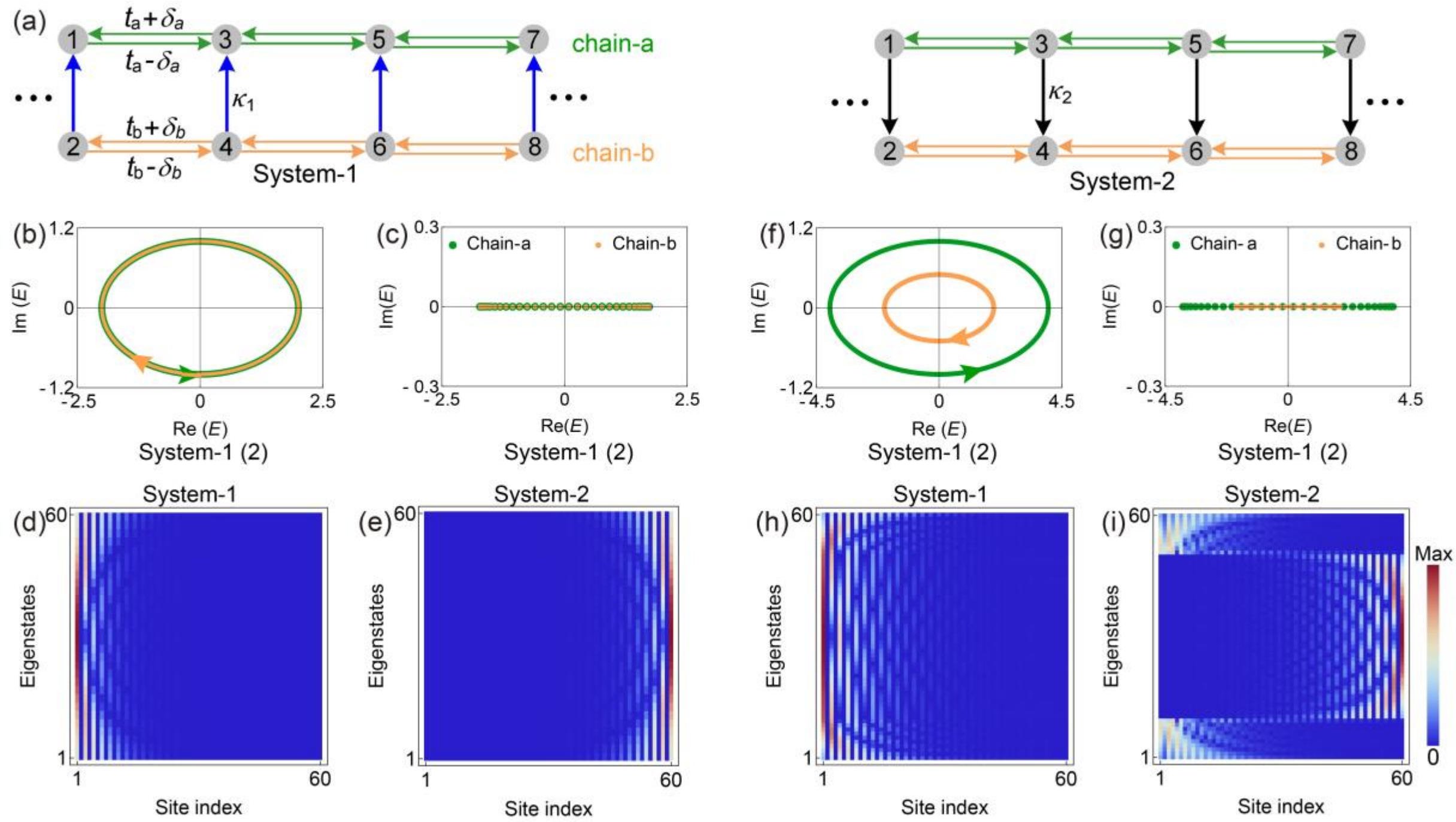

**Fig. S5 | ED produced in the model of two coupled Hatano-Nelson chains.** (a) The coupled HN models. (b and c) The PBC and OBC spectra of system-1(2) when the parameters are $t_a = t_b = 1$, $\delta_a = 0.5$, $\delta_b = -0.5$, and $\kappa_1 = \kappa_2 = 1$. (d and e) The corresponding OBC eigenstates of system-1 and system-2, respectively. (f and g) The PBC and OBC spectra of system-1(2) when the parameters are $t_a = 2$, $\delta_a = 0.5$, $t_b = 1$, $\delta_b = -0.25$, and $\kappa_1 = \kappa_2 = 1$. (h and i) The corresponding OBC eigenstates of system-1 and system-2, respectively. The green and orange colors in (b, c, f, g) represent the spectrum of chain-a and b, respectively.

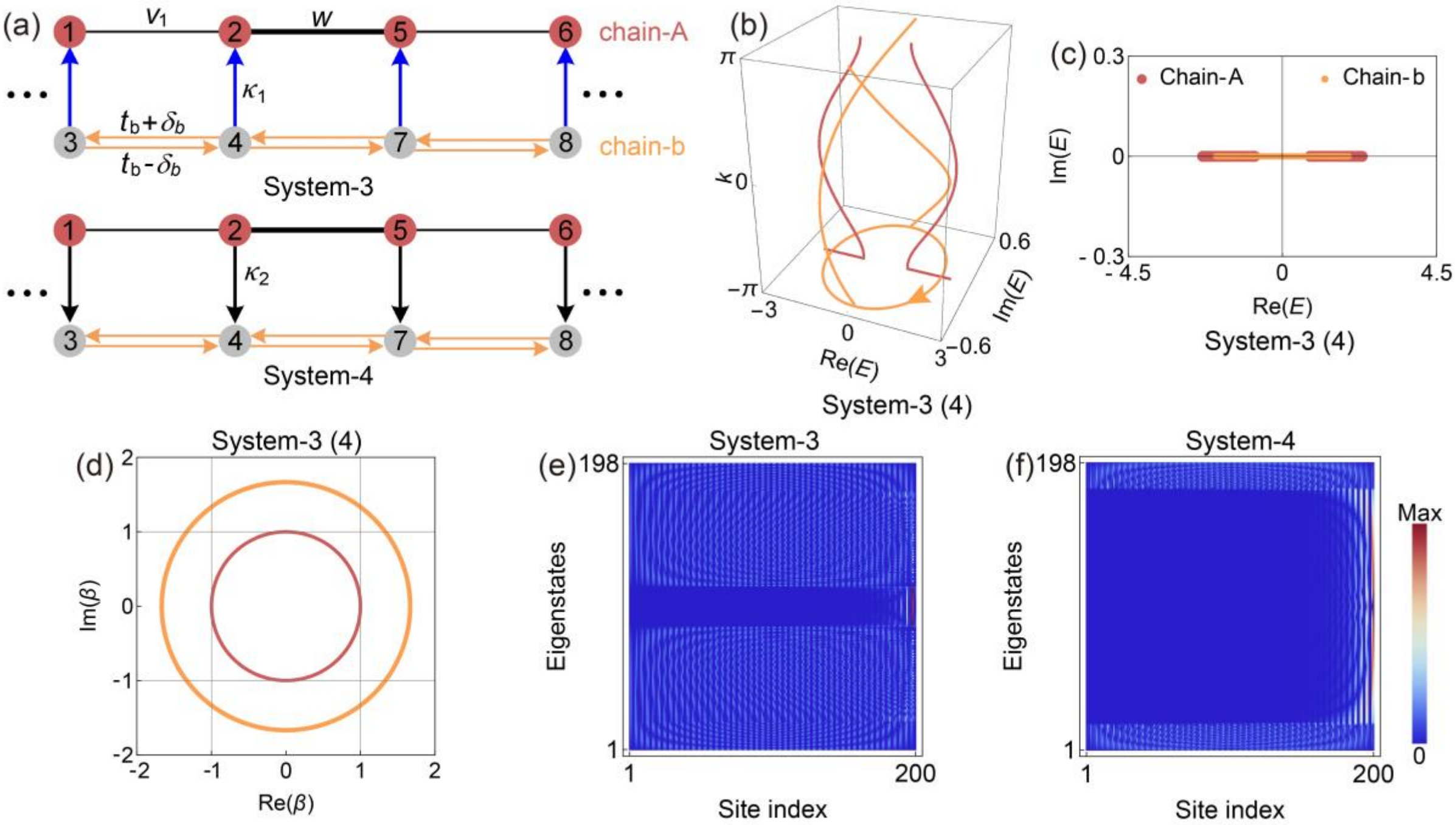

**Fig. S6 | ED induced effects in the model of two coupled different chains.** (a) The double-chain models consisted of a Hermitian SSH chain and a non-Hermitian HN chain. (b and c) PBC and OBC spectra of system-3 (4). (d) The GBZ of system-3 (4). (e and f) The OBC eigenstates of system-3 and system-4. The red and orange colors in (b to d) represent the

spectrum and GBZ of chain-A and chain-b, respectively. The parameters are $v_1 = -0.765$, $w = -1.565$, $t_b = 1$, $\delta_b = -0.25$, and $\kappa_1 = \kappa_2 = 10$. The zero-energy edge modes are omitted for clarity of presentation.

**4. Exactly solvable case**

In the special case of $t_a = -\delta_a$, we can exactly solve the eigenproblem of the additional model system-1. Hereafter we consider the following OBC Hamiltonian

$$\boldsymbol{h}_1 = \begin{pmatrix} \boldsymbol{h}_a & \boldsymbol{\kappa}_1 \\ 0 & \boldsymbol{h}_b \end{pmatrix}, \boldsymbol{h}_a = \begin{pmatrix} V & 0 & \cdots & 0 & 0 \\ J & V & \cdots & 0 & 0 \\ \vdots & \vdots & \ddots & \vdots & \vdots \\ 0 & 0 & \cdots & V & 0 \\ 0 & 0 & \cdots & J & V \end{pmatrix}, \boldsymbol{h}_b = \begin{pmatrix} 0 & tr & \cdots & 0 & 0 \\ tr^{-1} & 0 & \cdots & 0 & 0 \\ \vdots & \vdots & \ddots & \vdots & \vdots \\ 0 & 0 & \cdots & 0 & tr \\ 0 & 0 & \cdots & tr^{-1} & 0 \end{pmatrix}, \quad \text{(S1)}$$

which is just the system-1 (Fig. S5(a)) with specific parameters $J/2 = t_a = -\delta_a$, $t^2 = t_b^2 - \delta_b^2$, $r^2 = (t_b + \delta_b)/(t_b - \delta_b)$, and additional onsite potential $V \in \mathbb{C}$ in chain-a. Let $L$ be the length of each chain-a, b; Both $\boldsymbol{h}_a$ and $\boldsymbol{h}_b$ are $L \times L$ matrices. While $\boldsymbol{h}_a$ is non-diagonalizable and has a single eigenvalue $V$ with an eigenvector $(0, \dots, 0, 1)^{\mathrm{T}}$, $\boldsymbol{h}_b$ has $L$ distinct eigenvalues $\epsilon_q = 2t\cos q$, $q \in \left\{\frac{\pi}{L+1}, \frac{2\pi}{L+1}, \dots, \frac{L\pi}{L+1}\right\}$ with eigenvectors $u_q = (r^{-1}\sin q, r^{-2}\sin 2q, \dots, r^{-L}\sin Lq)^{\mathrm{T}}$. If $V \notin [-2t, 2t]$, the spectra of $\boldsymbol{h}_a$ and $\boldsymbol{h}_b$ do not overlap. Thus, the eigenvector $v_q$ of $\boldsymbol{h}_1$ with the eigenvalue $\epsilon_q$ is

$$v_q = \begin{pmatrix} \psi_q \\ u_q \end{pmatrix}, \quad \psi_q = -\boldsymbol{\kappa}_1 (\boldsymbol{h}_a - \epsilon_q)^{-1} u_q. \quad \text{(S2)}$$

Thanks to the specific form of $\boldsymbol{h}_a$, we know an analytic form of the inverse $(\boldsymbol{h}_a - \epsilon_q)^{-1}$ as

$$-(\boldsymbol{h}_a - \epsilon_q)^{-1} = \frac{1}{J} \begin{pmatrix} \frac{J}{\epsilon_q - V} & 0 & \cdots & 0 & 0 \\ \left(\frac{J}{\epsilon_q - V}\right)^2 & \frac{J}{\epsilon_q - V} & \cdots & 0 & 0 \\ \vdots & \vdots & \ddots & \vdots & \vdots \\ \left(\frac{J}{\epsilon_q - V}\right)^{L-1} & \left(\frac{J}{\epsilon_q - V}\right)^{L-2} & \cdots & \frac{J}{\epsilon_q - V} & 0 \\ \left(\frac{J}{\epsilon_q - V}\right)^{L} & \left(\frac{J}{\epsilon_q - V}\right)^{L-1} & \cdots & \left(\frac{J}{\epsilon_q - V}\right)^2 & \frac{J}{\epsilon_q - V} \end{pmatrix}. \quad \text{(S3)}$$

From this, we can write the $n$-th component of $\psi_q$ as

$$(\psi_q)_n = \frac{\kappa_1}{\epsilon_q - V} \frac{r^{-n}\left(\sin nq - \frac{rJ}{\epsilon_q - V}\sin(n+1)q\right) + r\left(\frac{J}{\epsilon_q - V}\right)^{n+1} \sin q}{1 - 2\frac{rJ}{\epsilon_q - V}\cos q + \left(\frac{rJ}{\epsilon_q - V}\right)^2}. \quad \text{(S4)}$$

It should be emphasized that the eigenvector $\psi_q$ on chain-a has two competing localization terms $r^{-n}$ and $\left(\frac{J}{\epsilon_q - V}\right)^{n+1}$.

Next, we normalize the eigenvector $v_q$ as

$$\tilde{v}_q = \frac{v_q}{\| v_q \|_F} = \frac{1}{\sqrt{\| \psi_q \|_F^2 + \| u_q \|_F^2}} \begin{pmatrix} \psi_q \\ u_q \end{pmatrix} \tag{S5}$$

and thereby we examine the infinite-volume behavior of the normalized eigenvector $\tilde{v}_q$ with fixed $q$. To do this, we evaluate the ration $\| \psi_q \|_F^2 / \| u_q \|_F^2$. After a long calculation, we have

$$\frac{\| \psi_q \|_F^2}{\| u_q \|_F^2} \overset{L\to\infty}{\to} \begin{cases} \infty & \text{if} \begin{cases} \left| \dfrac{rJ}{\epsilon_q - V} \right| \geq 1 \;\&\; 0 < r < 1 \\ \left| \dfrac{J}{\epsilon_q - V} \right| > 1 \;\&\; r = 1 \\ \left| \dfrac{J}{\epsilon_q - V} \right| \geq 1 \;\&\; r > 1. \end{cases} \\ \text{const.} & \text{otherwise.} \end{cases} \tag{S6}$$

Intuitively, this result can be interpreted as competition between two localization behaviors $r^{-n}$ and $\left(\frac{J}{\epsilon_q - V}\right)^{n+1}$. In case of $\left|\frac{J}{\epsilon_q - V}\right| \geq 1$ & $r > 1$, for instance, the term $\left(\frac{J}{\epsilon_q - V}\right)^{n+1}$ increases as one moves to the right of the chain while the term $r^{-n}$ is bounded, so $\| \psi_q \|_F$ containing both terms is larger than $\| u_q \|_F$ containing only the term $r^{-n}$. Accordingly, we find the normalized eigenvector exhibits an asymptotic behavior of

$$\tilde{v}_q \overset{L\to\infty}{\to} \begin{pmatrix} * \\ 0 \end{pmatrix} \quad \text{if} \begin{cases} \left| \dfrac{rJ}{\epsilon_q - V} \right| \geq 1 \;\&\; 0 < r < 1 \\ \left| \dfrac{J}{\epsilon_q - V} \right| > 1 \;\&\; r = 1 \\ \left| \dfrac{J}{\epsilon_q - V} \right| \geq 1 \;\&\; r > 1. \end{cases} \tag{S7}$$

In other words, the eigenvectors of $\boldsymbol{\hbar}_1$ get deficient in the Hilbert subspace $0 \oplus \mathbb{C}^L$ of chain-b after the infinite volume limit. It is noteworthy that such an asymptotic emergence of ED occurs only when $J \geq |\epsilon_q - J|$, which implies the eigenvalue $\epsilon_q$ is encircled by the PBC spectrum $\{V + Je^{ik} \mid k \in \mathbb{R}\}$ of $\boldsymbol{\hbar}_a$.

Lastly, we discuss about general criterion for asymptotic ED apart from the special case above. We just have to explore a condition for $\| (\boldsymbol{\hbar}_a - E)^{-1} \boldsymbol{\kappa}_1 u \|_F / \| u \|_F \overset{L\to\infty}{\to} \infty$, where $\boldsymbol{\hbar}_b u = Eu$. We take the largest singular value $s_L(E_0) = \| (\boldsymbol{\hbar}_a - E_0)^{-1} \|$ and the right-singular vector $w_L(E_0)$ of $(\boldsymbol{\hbar}_a - E_0)^{-1}$ for each $L$. Then we have

$$\frac{\| (\boldsymbol{\hbar}_a - E)^{-1} \boldsymbol{\kappa}_1 u \|_F}{\| u \|_F} \geq \frac{s_L(E_0) |w_L(E)^\dagger \boldsymbol{\kappa}_1 u|}{\| u \|_F}, \tag{S8}$$

which gives a sufficient condition for asymptotic ED as

$$\frac{\| (\boldsymbol{\hbar}_a - E)^{-1} \| \, |w_L(E)^\dagger \boldsymbol{\kappa}_1 u|}{\| u \|_F} \overset{L\to\infty}{\to} \infty. \tag{S9}$$

Because of $|w_L(E)^\dagger \boldsymbol{\kappa}_1 u| / \| u \|_F \leq \| \boldsymbol{\kappa}_1 \|$, this condition involves the divergence of the matrix norm $\| (\boldsymbol{\hbar}_a - E)^{-1} \|$ as long as $\| \boldsymbol{\kappa}_1 \|$ is uniformly bounded with respect to $L$. If the skin effect occurs for $\boldsymbol{\hbar}_a$, $\| (\boldsymbol{\hbar}_a - E)^{-1} \|$ for OBC Hamiltonian $\boldsymbol{\hbar}_a$ relates with the corresponding PBC spectrum. To see this, we focus on the pseudospectrum $\sigma_\epsilon(\boldsymbol{\hbar}_a)$, defined as the region of $z \in \mathbb{C}$ satisfying $\| (\boldsymbol{\hbar}_a - z)^{-1} \| > \epsilon^{-1}$. As known in the context of topological origin of skin effects, a reference energy $E_0$ at which the PBC spectrum has a nonzero winding number belongs to the limiting pseudospectrum $\lim_{\epsilon\to 0}\lim_{L\to\infty}\sigma_\epsilon(\boldsymbol{\mathscr{h}}_a)$ and therefore satisfies $\| (\boldsymbol{\mathscr{h}}_a - E_0)^{-1} \| \overset{L\to\infty}{\to} \infty$ [1]. Thus, a typical situation to realize Eq. (S9) is when the PBC spectrum of $\boldsymbol{\mathscr{h}}_a$ encircles an OBC eigenvalue $E$ of $\boldsymbol{\mathscr{h}}_b$ with nonzero winding induced by the skin effect which is consistent with the result in the specific model.

## 5. Additional experimental results

### 5.1. Green's function at ED

The steady-state response is obtained using the frequency-domain Green's function

$$\mathbf{G}(\omega) = (\omega\mathbf{I} - \mathbf{H})^{-1}, \tag{S10}$$

where $\omega$ is the angular frequency of excitation [2]. Let us derive the Green's function of a system under ED. Our double-chain Hamiltonian reads

$$\boldsymbol{\mathscr{h}} = \begin{pmatrix} \boldsymbol{\mathscr{h}}_A & \mathbf{0} \\ \mathbf{0} & \boldsymbol{\mathscr{h}}_B \end{pmatrix} + \begin{pmatrix} \mathbf{0} & \boldsymbol{\kappa}_1 \\ \boldsymbol{\kappa}_2 & \mathbf{0} \end{pmatrix} = \mathbf{H_0} + \mathbf{V}, \tag{S11}$$

Because we are concerned about ED, it is already assumed that $\boldsymbol{\mathscr{h}}_A$ and $\boldsymbol{\mathscr{h}}_B$ have the same spectrum. The Green's function of the system is

$$\begin{aligned}
\mathbf{G} &= (\omega\mathbf{I} - \boldsymbol{\mathscr{h}})^{-1} \\
&= [\omega\mathbf{I} - \mathbf{H_0} - \mathbf{V}]^{-1} \\
&= [(\omega\mathbf{I} - \mathbf{H_0}) - \mathbf{V}(\omega\mathbf{I} - \mathbf{H_0})^{-1}(\omega\mathbf{I} - \mathbf{H_0})]^{-1} \\
&= (\omega\mathbf{I} - \mathbf{H_0})^{-1}[\mathbf{I} - \mathbf{V}(\omega\mathbf{I} - \mathbf{H_0})^{-1}]^{-1} \\
&= \mathbf{G_0}[\mathbf{I} - \mathbf{V}\mathbf{G_0}]^{-1},
\end{aligned} \tag{S12}$$

where $\mathbf{G_0} \coloneqq (\omega\mathbf{I} - \mathbf{H_0})^{-1} = \begin{pmatrix} (\omega\mathbf{I} - \boldsymbol{\mathscr{h}}_A)^{-1} & \mathbf{0} \\ \mathbf{0} & (\omega\mathbf{I} - \boldsymbol{\mathscr{h}}_B)^{-1} \end{pmatrix} = \begin{pmatrix} \mathbf{G}_A & \mathbf{0} \\ \mathbf{0} & \mathbf{G}_B \end{pmatrix}$, and

$$[\mathbf{I} - \mathbf{V}\mathbf{G_0}]^{-1} = \begin{pmatrix} \mathbf{I} & \boldsymbol{\kappa}_1\mathbf{G}_B \\ \boldsymbol{\kappa}_2\mathbf{G}_A & \mathbf{I} \end{pmatrix} \begin{pmatrix} (\mathbf{I} - \boldsymbol{\kappa}_1\mathbf{G}_B\boldsymbol{\kappa}_2\mathbf{G}_A)^{-1} & \mathbf{0} \\ \mathbf{0} & (\mathbf{I} - \boldsymbol{\kappa}_2\mathbf{G}_A\boldsymbol{\kappa}_1\mathbf{G}_B)^{-1} \end{pmatrix}.$$

Therefore, the Green's function is defined as

$$\begin{aligned}
\mathbf{G} &= \mathbf{G_0}[\mathbf{I} - \mathbf{V}\mathbf{G_0}]^{-1} \\
&= \begin{pmatrix} \mathbf{G}_A & \mathbf{G}_A\boldsymbol{\kappa}_1\mathbf{G}_B \\ \mathbf{G}_B\boldsymbol{\kappa}_2\mathbf{G}_A & \mathbf{G}_\mathbf{B} \end{pmatrix} \begin{pmatrix} (\mathbf{I} - \boldsymbol{\kappa}_1\mathbf{G}_B\boldsymbol{\kappa}_2\mathbf{G}_A)^{-1} & \mathbf{0} \\ \mathbf{0} & (\mathbf{I} - \boldsymbol{\kappa}_2\mathbf{G}_A\boldsymbol{\kappa}_1\mathbf{G}_B)^{-1} \end{pmatrix}.
\end{aligned} \tag{S13}$$

Consider system-I in the main text, the ED is reached when $\boldsymbol{\kappa}_2 = \mathbf{0}$, so Eq. (S13) reduces to

$$\mathbf{G}_{ED} = \begin{pmatrix} \mathbf{G}_A & \mathbf{G}_A \boldsymbol{\kappa}_1 \mathbf{G}_B \\ \mathbf{0} & \mathbf{G}_B \end{pmatrix}. \tag{S14}$$

Equation (S14) is the key to understanding the steady-state response measured in our experiment. When a source is placed at chain-A, the steady-state response is determined by the first column of Eq. (S14), which only has $\mathbf{G}_A$. This is well-expected because the eigenspace $\mathcal{E}(\boldsymbol{\hbar}_B)$ is "missing" at the ED, and a source at $\boldsymbol{\hbar}_A$ overlaps only with $\mathcal{E}(\boldsymbol{\hbar}_A)$. But a source incident at chain-B produces responses given by the second column of Eq. (S14), which indicates that both chains are excited and the effect of the defective eigenspace manifests through $\mathbf{G}_B$ in both chains.

### 5.2. Experimental results with different $v_1$

We have performed additional experiments to validate the eigenvalue spectra of the spectral condition for ED, i.e., $\eta(\boldsymbol{\hbar}_A) = \eta(\boldsymbol{\hbar}_B)$. We tuned the intra-cell hopping $v_1$ by changing the corresponding spring constant, which changes $\eta(\boldsymbol{\hbar}_A)$. We then measure the frequency responses of chain-A by exciting at site 14 chain-A (site 16 at chain-B) with a chirp signal that covers the frequency regime of all bands. (It also induces a slight shift in the natural frequency of the oscillators in chain-A. The natural frequency of the oscillators in chain-B is also fine-tuned to match that of chain-A.) The measured results, together with the theoretical ones, are shown in Fig. S7. When $v_1 < v_e$ ($v_e \approx -0.760$), the OBC spectrum of chain-B spans a wider frequency range than that of chain-A, and the opposite is seen for $v_1 > v_e$. When $v_1$ is close to $v_e$, the eigenvalue spectra of chain-A and B, as well as their frequency responses, are nearly identical. These characteristics are successfully observed.

The corresponding steady-state responses were measured and are shown in Fig. S8 and Fig. S9. In these measurements, system-I (II) was excited at chain-B on site 8 (chain-A on site 6) using sinusoidal signals at 14 different frequencies.

For $v_1 < v_e$, both extended states and skin modes are observed in system-I. Skin modes are observed for an excitation frequency at band edges, whereas delocalized responses indicating the dominance of extended modes are seen inside the bands. These features are consistent with the characteristics of the OBC eigenstates [Fig. S7]. In system-II, the response is dominated by skin modes on chain-B.

For $v_1 = -0.765$, which is very close to the ED, the spectra of chain-A and B are identical, hence the steady-state responses are dominated by the ED.

When $v_1 > v_e$, the steady-state responses in system-I are dominated by the delocalized responses across the entire excitation-frequency range. In system-II, when the excitation frequency lies in the middle of the energy bands, the skin-mode response on chain-B is the dominant feature. However, as the frequency approaches the band edges, the extended response is also observed. Further increase of $v_1$ produces wider bands in chain-A, such that more extended states are seen, which is also observed in our experiments (compare the rows for $v_1 = -0.970, -1.194$ in Fig. S8).

To better summarize the results, we sum the magnitudes of the responses of the same type, i.e., extended or localized, over all measured frequencies. The results are plotted in Fig. S9. For $v_1 < v_e$, both extended states and skin modes are observed in system-I, whereas in system-II this occurs when $v_1 > v_e$. The states within the overlapping parts of their OBC spectra remain dominated by ED. These results also agree well with the responses obtained by Green's function calculation.

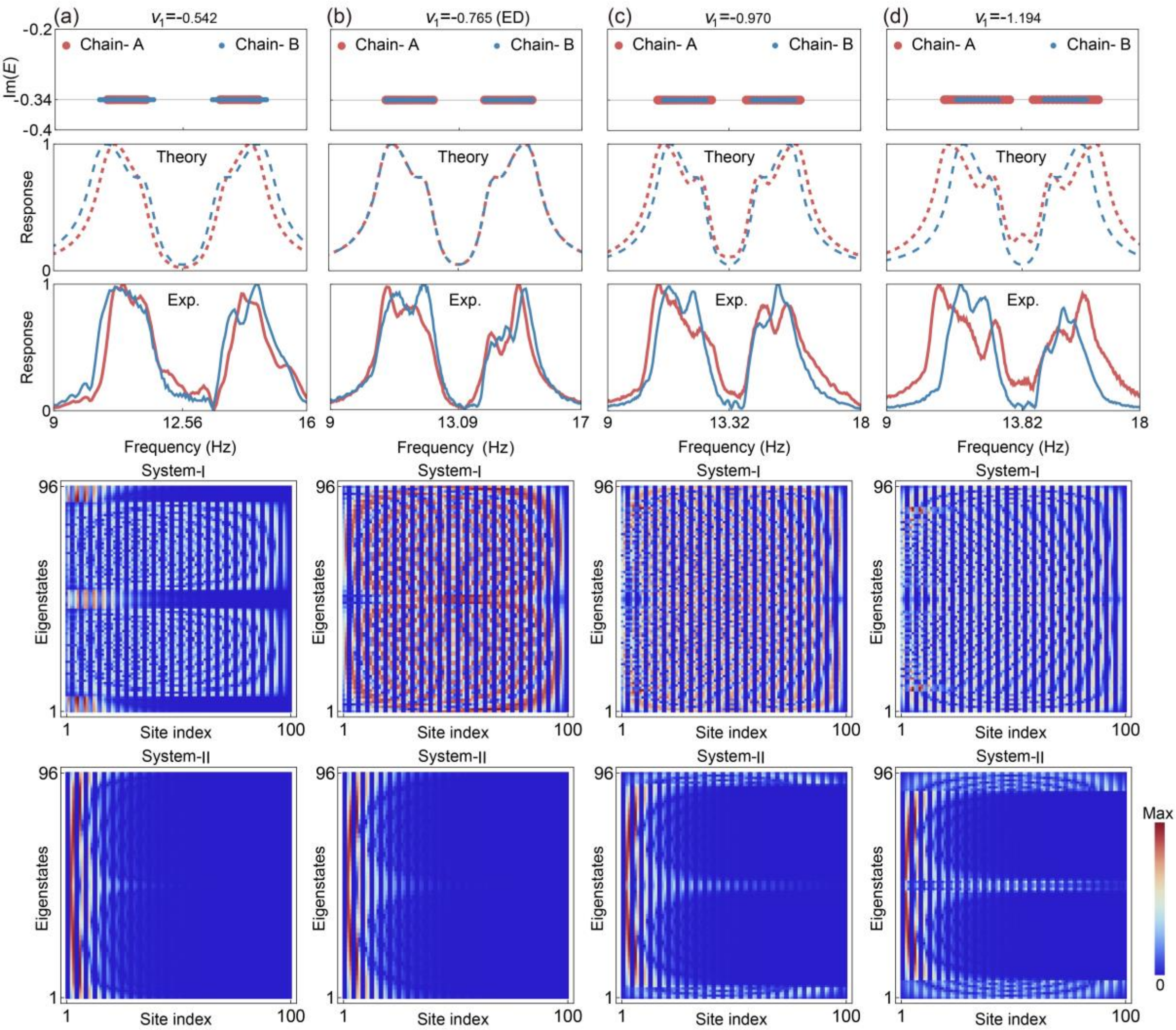

**Fig. S7 | The eigenvalues, frequency responses, and eigenstates for systems with different values of $v_1$.** (a) $v_1 = -0.542$. (b) $v_1 = -0.765$ (close to ED). (c) $v_1 = -0.970$. (d) $v_1 = -1.194$. The first two rows show theoretically calculated OBC eigenvalue spectra and frequency responses. The third row is the experimentally measured frequency responses, the horizontal axis of which reveals slight shifts in the oscillators' natural frequencies. The fourth and fifth rows show theoretical eigenstates for system-I and II.

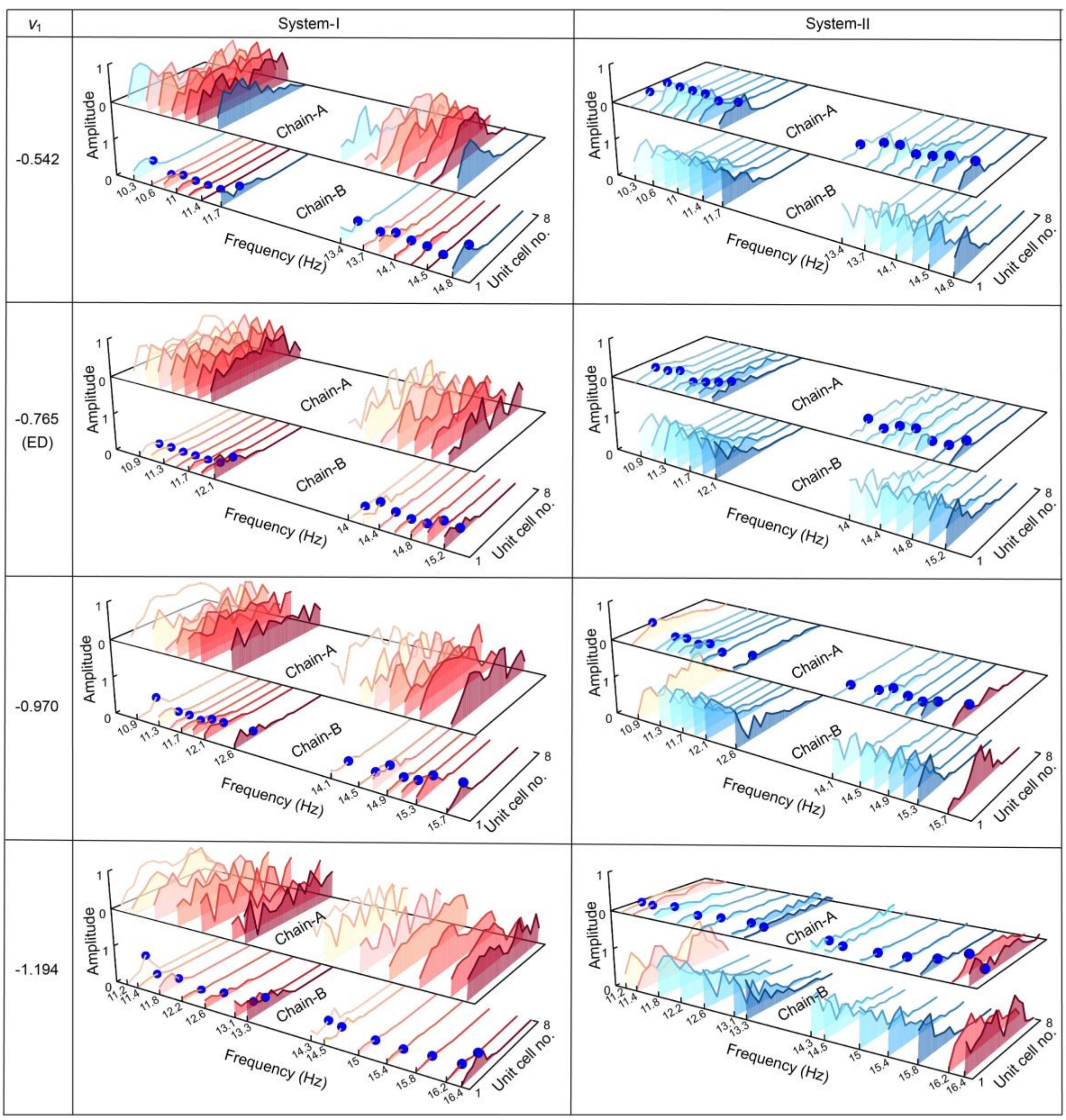

**Fig. S8 | Experimental results of the steady-state response of system-I and II.** The two systems are excited by sinusoidal signals at 14 different frequencies. The shaded red and blue lines distinguish the responses of extended states and skin modes. The position of the source is fixed at chain-B on site 8 (chain-A on site 6) in the measurement of system-I (II), as indicated by the blue dots.

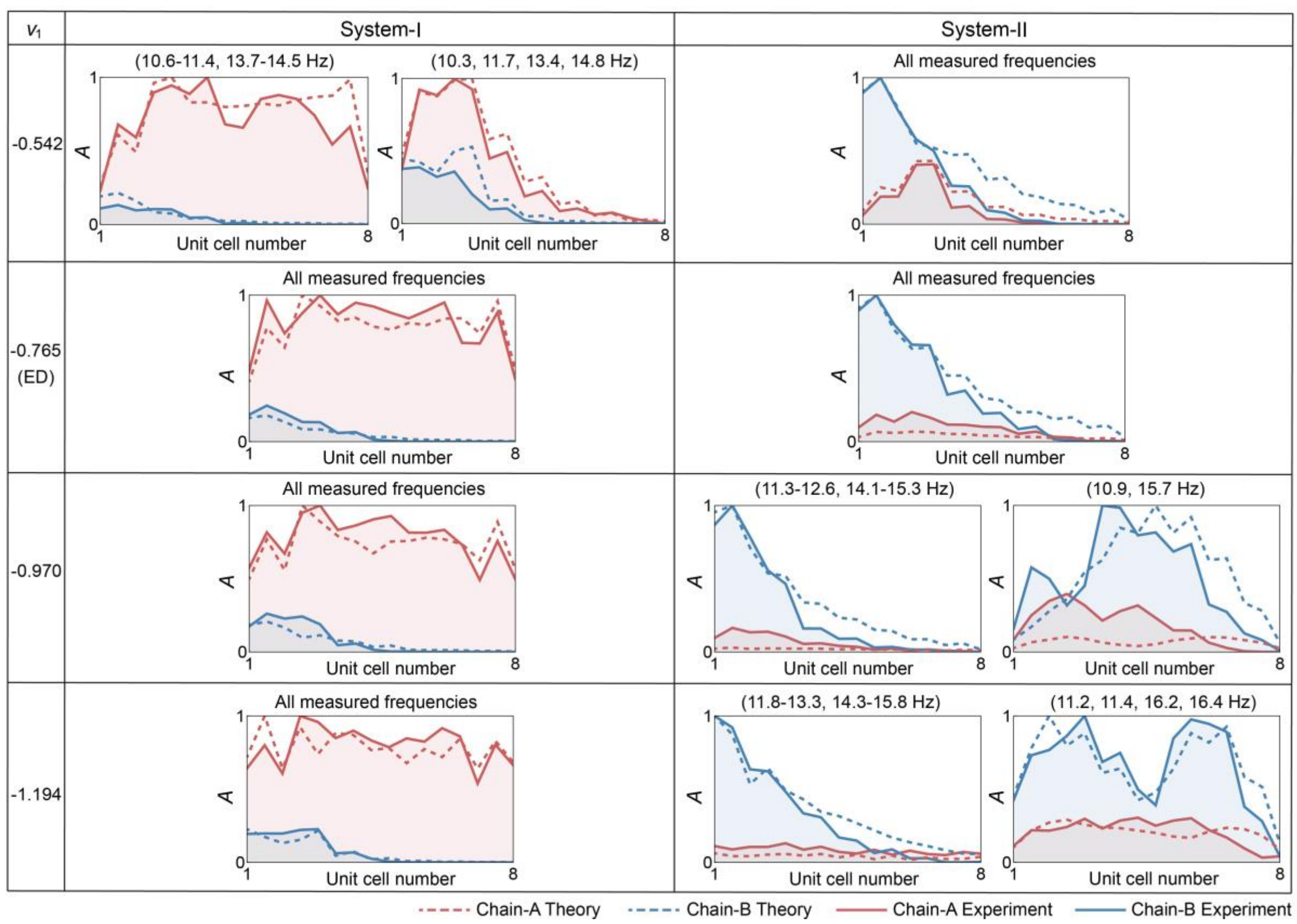

**Fig. S9 | The summed delocalized and skin-mode responses over all measured frequencies in Fig. S8.** The amplitude at each site is defined as $A_n = \sum_i \theta_n(f_i)$, where $n$ is the site index, and $i$ indexes the measured frequency $f$ with delocalized or skin-mode responses. The frequencies summed over are marked in each figure. The dashed red (blue) and shaded solid red (blue) lines plot computed and measured results of chain-A (chain-B), respectively.

### 5.3. Steady-state responses with $\eta(\hbar_B)$ offset

In Fig. S10(a-c), it is seen that the steady-state response of system-I when a single-frequency excitation at different frequencies is applied to chain-A. As illustrated in Fig. S10(a, b), the wavepacket propagates to both edges of the system, demonstrating the properties of bulk modes. When the excitation frequency is outside the eigenfrequency range of chain-A, the wavepacket decays significantly as it moves away from the source [Fig. S10(c)]. For system-II, when an excitation with a frequency lower than the bands of chain-B is applied to chain-B, the wave is seen to attenuate [Fig. S10(d)]. When the excitation frequency is inside the bands of chain-B, skin modes are seen [Fig. S10(e, f)].

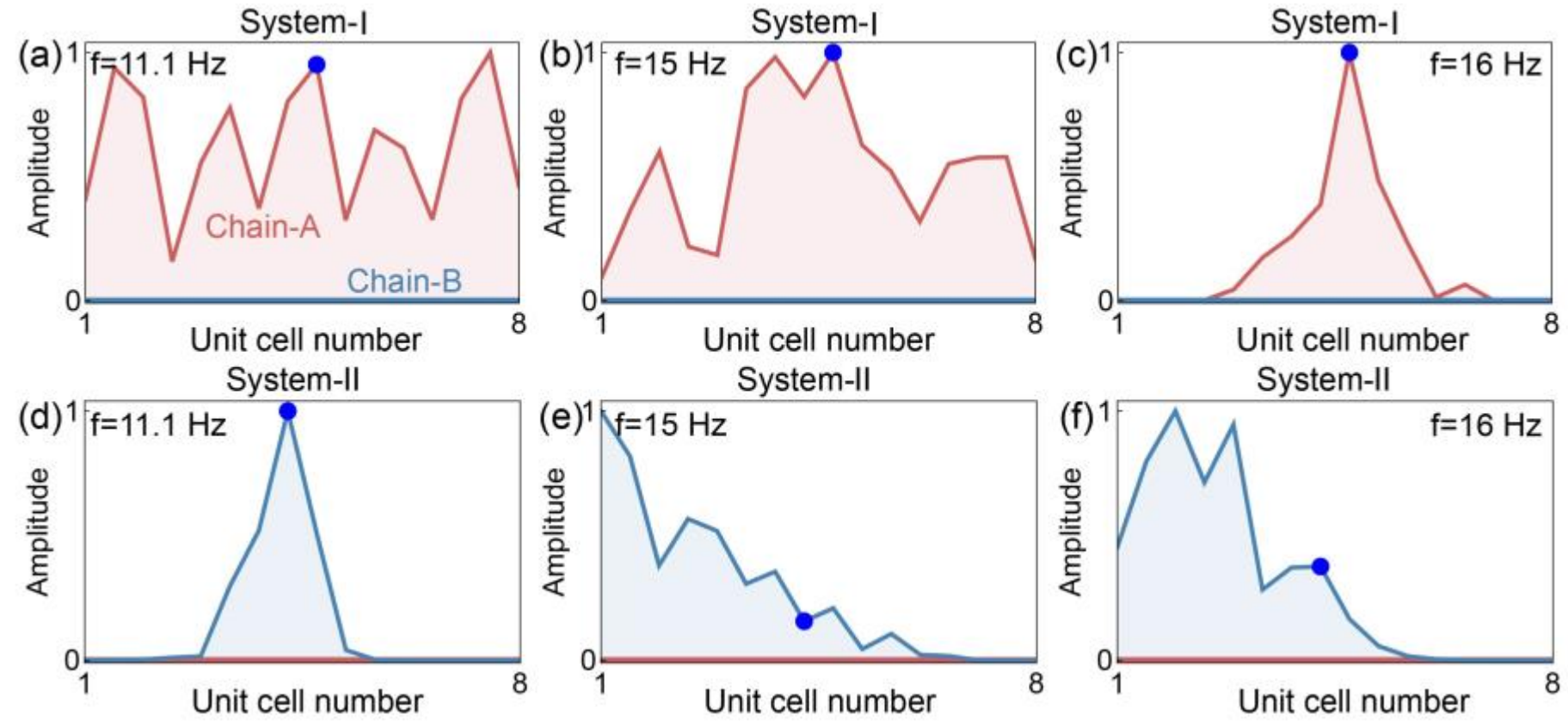

**Fig. S10 | Steady-state response of system-I (system-II) as the source is at chain-A (chain-B).** (a-c) Experimentally measured steady-state response of system-I when the excitation is fixed at chain-A on site 17 (indicated by the blue dots). (d-f) Experimental results for system-II, with the excitation is placed at chain-B on site 16 (indicated by the blue dots). The red and blue lines plot the responses of chain-A and B, respectively. The natural frequencies of chain-A and chain-B are $f_A = 13.09$, $f_B = 14.12$. All other parameters remain the same.

## 6. Effects of perturbation on dynamics

To explore the effects of perturbations on the dynamic behaviors of system-I and II, we introduced three distinct sets of perturbations to each system and computed the corresponding dynamical behaviors, as shown in Fig. S11.

In System-I ($\kappa_1 \neq 0, \kappa_2 = 0$), as $\kappa_2$ is increased from zero to $10^{-7}$, a slight temporal growth of the wavepacket is already observed when the excitation is applied to chain-A. But the skin-effect amplified propagation remains clearly seen when the excitation is at chain-B. With further increase in $\kappa_2$, the location of excitation is no longer consequential. Notably, because the OBC spectrum does not lie on the real axis, the wavefunctions are eventually dominated by the exponential growth caused by positive imaginary parts in the eigenenergy.

Similar behaviors are also observed in System-II ($\kappa_2 \neq 0, \kappa_1 = 0$): small perturbation in $\kappa_1$ does not markedly alter the system dynamics. But a larger $\kappa_1$ drives the system away from the ED and the dynamics disregards the location of excitation, and the long-time evolution is dominated by the imaginary eigenenergy. Also, the difference between system-I and II can no longer be distinguished.

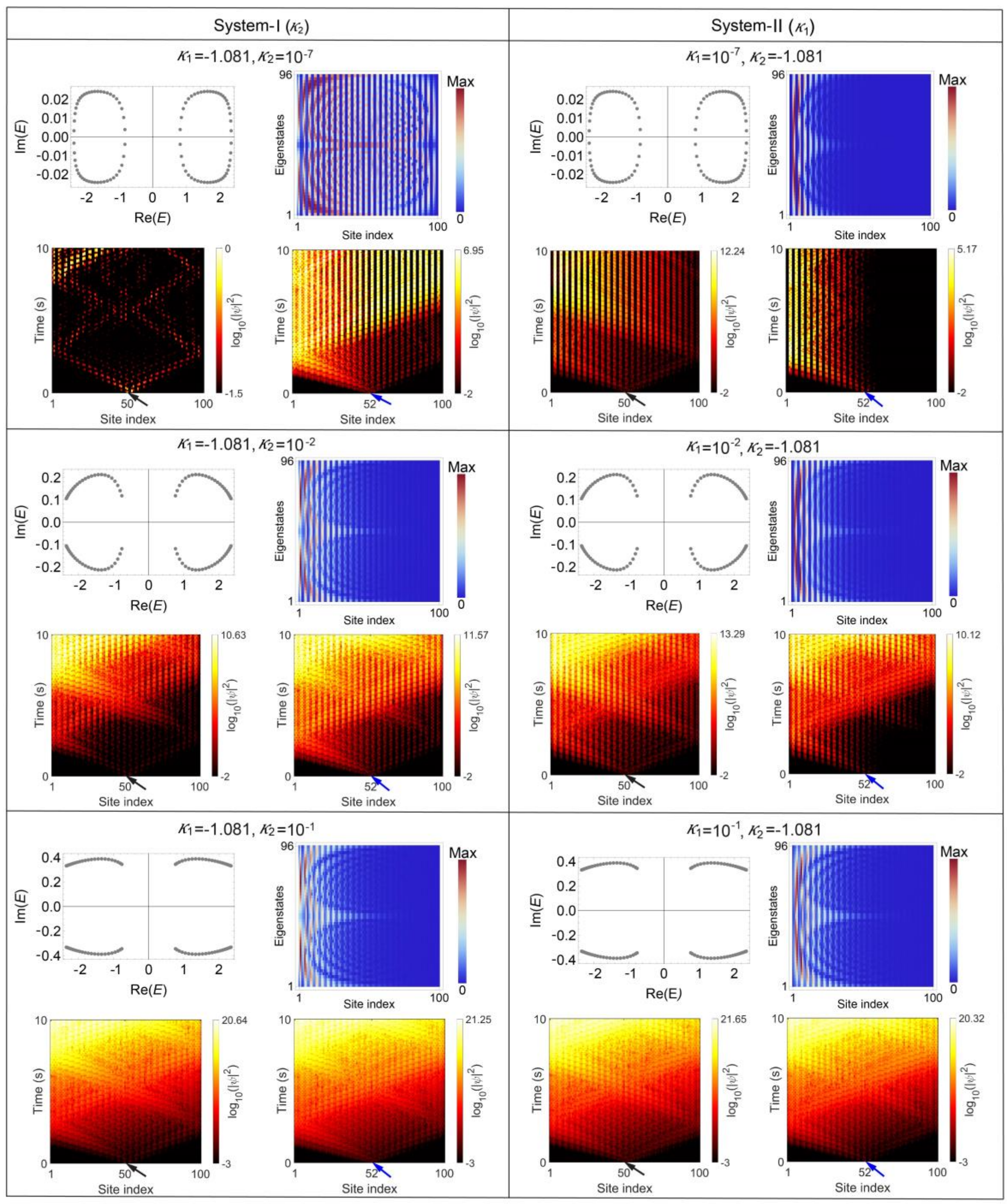

**Fig. S11 | Dynamic behaviors of system-I (system-II) with perturbations.** The perturbations are $\kappa_2(\kappa_1) = 10^{-7}, 10^{-2}, 10^{-1}$ from the first to the third row. The initial wavepacket is injected at the center of chain-A (marked by the black arrow), and chain-B (marked by the blue arrow), respectively. The parameters are the same as Fig.2 in the main text.